\documentclass[aps,prd,twocolumn,groupedaddress]{revtex4-2}

\usepackage{CJK}

\usepackage{natbib}

\usepackage{graphicx}  
\usepackage{dcolumn}   
\usepackage{bm}        
\usepackage{amssymb}   
\usepackage{amsmath}
\usepackage{slashed}
\usepackage{xspace}
\usepackage{epstopdf}
\usepackage{color}
\usepackage{bbm}
\usepackage{cancel}
\usepackage{ulem} 

\usepackage{times}

\newcommand{\s}{$S$}
\newcommand{\sdm}{$S$DM}
\newcommand{\sbar}{$\bar{S}$}
\newcommand{\gsbb}{$ \tilde{g}$}
\newcommand{\gssv}{$ g_{SSV}$}
\newcommand{\asn}{$ \alpha_{SN}$}

\newcommand{\n}{$N$}

\newcommand{\mnras}{MNRAS}
\newcommand{\aap}{Astronomy \& Astrophys.}

\newcommand{\li}{$^7{\rm Li}$}
\newcommand{\Be}{$^7{\rm Be}$}

\newcommand{\He}{$^4{\rm He}$}

\newcommand{\be}{\begin{equation}}
\newcommand{\ee}{\end{equation}}
\newcommand{\ba}{\begin{align}}
\newcommand{\ea}{\end{align}}

\newcommand{\ODMOb}{$\Omega_{DM}/\Omega_b$}

\begin{document}
\begin{CJK*}{UTF8}{gbsn}

\title{Dark Matter Particle in QCD}


\author{Glennys R. Farrar} 
\author{\small Zihui Wang (王子汇)} 
\author{\small Xingchen Xu (许星辰)}
\affiliation{Center for Cosmology and Particle Physics,
Department of Physics,
New York University, NY, NY 10003, USA}
 \email[]{gf25@nyu.edu}

\date{\today}

\begin{abstract}
We report on the possibility that the Dark Matter particle is a stable, neutral, as-yet-undiscovered hadron in the standard model.  We show that the existence of a compact color-flavor-spin singlet $uuddss$ (Sexaquark, \s) with mass of order $2 \, m_p$ is compatible with current knowledge and that, if it exists, the \s\ is a very attractive DM candidate.  The \s\ interacts with baryons primarily via a Yukawa interaction of coupling strength \asn, mediated by exchange of the flavor-singlet superposition of the $\omega$ and $\phi$ vector mesons, denoted $V$, having mass $\approx 1$ GeV.  We emphasize the need to distinguish between \s-nucleon scattering amplitudes
which are of a hadronic scale, and \s\ breakup amplitudes 
which are dynamically suppressed and many orders of magnitude smaller, akin to the weak interaction level.   We use SNOlab and other data to obtain the most stringent constraints on the effective vertex for breakup, \gsbb, from the stability of DM and nuclei.  The relic abundance of \s\ Dark Matter (\sdm) is established when the Universe transitions from the quark-gluon plasma to the hadronic phase at $\approx 150$ MeV and is in remarkable agreement with the observed $\Omega_{DM}/\Omega_b = 5.3 \pm 0.1$; this is a no-free-parameters result because the relevant parameters are known from QCD.   Survival of this relic abundance to low temperature requires $\tilde{g} \lesssim 2 \times 10^{-6}$, comfortably compatible with theory expectations and observational bounds.   To analyze bounds on \sdm\ we must solve the Schroedinger equation to determine the cross section, $\sigma_{A}$, for \s\ scattering on nucleus $A$.  Depending on  \asn, the true cross section 
can be orders of magnitude larger or smaller than given by Born approximation;  this requires a reanalysis of observational limits.  We use direct detection experiments and cosmological constraints to determine the allowed region of \asn\ for the mass range relevant to \sdm.  
If the \s-nucleon interaction is attractive and strong enough, DM-nucleus bound states will form.  For a range of allowed values of \asn, we predict exotic nuclear isotopes at a detectable level with mass offset $\approx 2$ amu.  Dedicated study of this mass-offset range, for a wide range of elements, is warranted.    
We argue that the neutron-star equation of state and SN1987a cooling are not constraining at this time, but could become so in the future when better understood.   Finally, we discuss strategies for detecting the sexaquark in accelerator experiments.  This is surprisingly difficult and experiments to date would not have discovered it.  The most promising approaches we identify are to search for a long-interaction-length neutral particle component in the central region of relativistic heavy ion collisions or using a beam-dump setup, and to search for evidence of missing particle production characterized by unbalanced baryon number and strangeness using Belle-II or possibly GLUEX at J-Lab.
\end{abstract}


\maketitle
 \end{CJK*}

\section{\label{sec:intro} Introduction}

A successful model for dark matter (DM) must predict the observed relic DM density and ideally also provide a natural explanation for the observed DM to baryon ratio, \ODMOb = 5.3$\pm 0.1$~\cite{PlanckCosmo15}.  It must be compatible with cosmological and astrophysical constraints on structure formation and DM interactions and not alter or interfere with the successful predictions of primordial nucleosynthesis. The DM interactions with normal matter must also satisfy direct detection bounds and constraints from laboratory and geophysical experiments, and must be compatible with observed properties of galaxies, neutron stars, white dwarfs, supernovae, and other astrophysical objects.

We show here that the sexaquark \s\ -- a conjectured neutral, color-flavor-spin-singlet bound state of six light quarks $u  u  d d s s$ with mass $ m_S \approx 2 m_p$ -- satisfies or is compatible with all of the above criteria given present limits to our understanding.  For $m_S < m_D + m_e$ the \s\ is absolutely stable and for $m_S \lesssim 2$ GeV its lifetime is greater than the age of the Universe.  The potential existence of this state and its compatibility with accelerator experiments was discussed in \cite{fS17}, where it was called sexaquark, adopting the Latinate prefix to distinguish it from the relatively loosely bound H-dibaryon proposed by Jaffe \cite{jaffe:H} and the term hexaquark which is a generic term for a 6-quark or $(q \bar{q})^3$ state; \s\ is also a reminder that it is a strange, scalar, flavor singlet.  

The relic abundance of sexaquark DM (\sdm) follows from general arguments of statistical physics and known standard model parameters -- the quark masses and the temperature of the transition from quark-gluon to hadronic phases -- and is predicted to be  \ODMOb\ $\approx 5$ \cite{fudsDM18}, in remarkable agreement with the observed value \ODMOb\ $= 5.3 \pm 0.1$ \cite{pdg18}.  Preservation of this abundance ratio as the Universe cools requires that the rate for breaking up \s's in hadronic collisions be less than the expansion rate of the Universe.  This condition is satisfied if the effective Yukawa vertex for breakup $\tilde{g} \lesssim$ few $10^{-6}$ \cite{fudsDM18}; this small value naturally follows from the low probability of fluctuation between di-baryon and sexaquark configurations \cite{fzNuc03} as discussed further in Sec. \ref{sec:bu}.  
For future reference, the mean number density of \sdm\ is about 2.5 times that of baryons, since $m_S \approx 2 m_p$.  

The organization of this paper is the following.  In Sec.~\ref{SS} we briefly review the particle physics of the proposed \s.  Then in Sec.~\ref{sec:DMabun} we give the DM abundance analysis predicting  \ODMOb\ $ \approx 5$ in the \sdm\ scenario at freezeout. In Sec. \ref{sec:bu} we discuss theoretical estimates of the breakup amplitude \gsbb, and provide improved observational limits as a function of sexaquark mass based on deuteron and sexaquark stability and other constraints.   With these basics in place, we proceed to the other requirements of a successful DM model starting with \sdm-matter interactions.  
The primary coupling of the \s\ to other hadrons is through exchange of the flavor-singlet superposition of the nonet vector mesons, whose mass is of order 1 GeV.  The resulting Yukawa interaction between \sdm\ and baryons is non-perturbative over important parts of the relevant parameter range, so the Schroedinger equation must be solved numerically to find the cross sections; this is discussed in Sec. \ref{sec:DMints}.  
Using the exact non-perturbative treatment, in Sec.~\ref{sec:DDconstraints} we derive the constraints on the Yukawa interaction strength $\alpha_{SN} $ implied by direct detection experiments, cosmology and astrophysics.  The limits are drastically different than would be deduced using Born approximation.   In Sec.~\ref{sec:SIDM} we investigate the self-interactions of \sdm\ and show that the maximum SIDM cross section is $\sigma/m \approx 0.2$ which is lower than generally considered astrophysically useful.  Next, in Sec.~\ref{sec:bbnhybrids}, we discuss the possible formation of exotic isotopes in which a sexaquark binds to a nucleus, and the constraints which can be placed on the parameter space from those considerations.  We find that for an interesting range of currently-allowed parameters there can be a sufficient density of exotic isotopes to be detectable, albeit requiring a new, dedicated search because previous limits are not sensitive to the $\approx 2$ amu splitting.   Section~\ref{sec:searches} discusses ways to search for sexaquarks in accelerator experiments.   Sec.~\ref{sec:Sum} gives a concise summary of the results of the paper, and we close with conclusions in Sec~\ref{sec:Conc}.   The Supplemental Material provide additional information on secondary topics.

\section{\label{SS} Stable Sexaquark Hypothesis}
The stable sexaquark hypothesis \cite{fS17} postulates that the Q=0, B=+2, $uuddss$ flavor-singlet scalar bound state (denoted \s) is  stable.  The \s\ is absolutely stable if $m_S  \le m_D + m_e = 1876.12$ MeV.  A somewhat higher mass can also effectively be stable, because up to $m_S = m_p+m_e + m_\Lambda = 2054.5 $ MeV the \s\ decays through a doubly-weak interaction and its lifetime may be longer than the age of the Universe \cite{fzNuc03}.  Both cases are called ``stable" below for conciseness.   The $S$ cannot be too light, or nuclei would decay.  These constraints are discussed in greater detail in Sec.~\ref{sec:bu}.

The stable sexaquark hypothesis is motivated by the unique symmetry of the $uuddss$ ground state.  Models designed to fit known hadrons cannot be trusted to reliably describe it because Fermi statistics prevents mesons and baryons from enjoying the triply-singlet configuration (in color, flavor, spin) accessible to $uuddss$.   Hyperfine attraction is strongest in singlet configurations, c.f., the Most-Attractive-Channel hypothesis \cite{MAC}, so binding is maximal in the sexaquark channel.   

Lattice studies are not yet capable of determining the mass spectrum of the $uuddss$ system.  A nearly unbound state is predicted by HAL-QCD~\cite{HALQCD18}, a lattice-inspired approach to modeling the physical light quark mass limit which however has been criticized~\cite{YamazakiKLatBS17,DavoudiLatticeNuc17}.  The NPLQCD group using 850 MeV $u,d,s$ quarks found 80 MeV binding energy in the H-dibaryon channel.  Rigorous lattice treatment of a 6-quark system, for physical quark masses, large volume and statistical sensitivity adequate to be sensitive to the presence of multiple states is extremely challenging and may be many years away.   It should be emphasized that there is no incompatibility between the existence of a deeply bound stable \s\ and a loosely bound  di-$\Lambda$ molecule analogous to the deuteron, for which there may be hints in the recent femtoscopy study by ALICE\cite{ALICE_LamLam19}.  The presence of such a loosely bound di-$\Lambda$ would complicate lattice QCD attempts to isolate an orthogonal, deeply bound state.

If it exists, the \s\ should be much more compact and weakly coupled than normal hadrons due to being a flavor singlet and thus not coupling to pions.  Baryons ($r_N = 0.9$ fm) are much larger than their Compton wavelength ($\lambda_N = 0.2$ fm), which can be attributed to baryons coupling to pions ($\lambda_\pi = 1.4$ fm).   Estimating 
\be
\label{rSb}
r_S = \lambda_S + b \lambda_{M1}
\ee
with $0\leq b < 0.45$ by analogy with baryons, where $M1$ is the lightest well-coupled flavor singlet meson, presumably the flavor-singlet combination of $\omega-\phi$ with $m_{M1} \sim 800-1000$ MeV, gives $r_S = 0.1-0.3$ fm.   

The disparate size of \s\ and baryons means amplitudes for breakup and formation reactions involving overlap of \s\ and two baryons, are very suppressed;  see Sec. \ref{sec:bu} below for more details.   Amplitudes for \s-nucleon scattering should be smaller than hadronic scattering amplitudes like $NN,~ \pi N$, etc., due to the absence of pion exchange, but this is less dramatic phenomenologically because the flavor singlet vector meson contribution remains.

Initial searches for a $uuddss$ bound state were stimulated by Jaffe's MIT bag model estimate of 2150 MeV \cite{jaffe:H} for a state he called H-dibaryon.  With a mass below $2 m_\Lambda = 2230$ MeV, the state is strong-interaction stable and was almost universally assumed to have a typical weak lifetime $\gtrsim 10^{-10}$ s as a result of expecting $m_H > m_p + m_\Lambda = 2054$ MeV.  Additionally, the H was envisaged structurally as a loosely-bound di-$\Lambda$, readily formed in hypernuclei, e.g., \cite{Baym+JaffeCygX3_85}.   Dozens of experiments were performed attempting to find an H-dibaryon, and seem to exclude the original proposal of a di-$\Lambda$ bound by $\mathcal{O}($100 MeV).  

A careful re-examination of the experimental situation by one of us (GRF) showed that no experiment to date would have detected a compact, stable \s~\cite{fS17} .  Experiments either required $m_H > 2$ GeV, or searched for a signal in the invariant mass of decay products such as $\Lambda p \pi^- $, or implicitly assumed a dibaryon spatial configuration comparable to a deuteron or nucleon so its interactions and production was expected to be comparable to ordinary hadrons; see \cite{fS17} for further discussion.  

If a stable \s\ exists, it could be the Dark Matter particle.  Limits on DM-baryon interactions from direct detection experiments, the CMB power spectrum and the indirect limits of \cite{nfm18} from HST orbital decay and evaporation of liquid cryogens, and thermal conductivity of the Earth, are discussed in Sec. \ref{sec:DMints} below; these limits prove to be only mildly restrictive on the natural parameter space. 

\section{Dark Matter relic abundance}
\label{sec:DMabun}
\subsection{QCD phase transition}
\label{QGPtran}

At high temperature, the QCD sector consists of a plasma of massless gluons, nearly massless $u, \bar{u},d,\bar{d}$ quarks and somewhat heavier $s,\bar{s}$ quarks.   At low temperature, the QGP is replaced by the chiral-symmetry-broken, color-confined phase in which baryons are heavy and pseudoscalar mesons are light.   Lattice QCD calculations show that the transition between the QGP and the low temperature hadronic phase is a cross-over centered on 155 MeV \cite{hotQCD14} rather than a true phase transition.  As the temperature drops from 170 MeV to 140 MeV, the quark and gluon condensates responsible for hadron masses and color confinement increase; at the same time it becomes more favorable energetically for $q \bar{q}$'s and $qqq$ to combine into color singlet mesons and baryons.   Typical intra-$q, \bar{q},g$ separations are $\mathcal{O}(1$ fm) for $T \approx 150$ MeV.  The age of the Universe in this epoch is 
$
t_{\rm Univ} = 7.3 \times10^{-5} (100 {\rm\, MeV}/T)^2 \, {\rm sec}, 
$
whereas the timescale for hadronic interactions is $\mathcal{O}(10^{-23}$s).

The equilibrium number density of each fermion species as a function of temperature is given by 
\be
\label{nq}
n(m,T) = \frac{g}{2 \pi^2} \int_m^\infty \frac{E \sqrt{E^2 - m^2}}{e^{(E \mp \mu)/T} + 1}\, dE, 
\ee
where $g=6$ is the number of color-spin degrees of freedom per $q$ and $\bar{q}$ flavor) and $\mu$ is the chemical potential.   

The quark masses are accurately known from the hadron spectrum in lattice QCD \cite{bazazov+QuarkMasses18}: $m_u = 2.118(38)$ MeV, $m_d = 4.690(54)$ MeV and $m_s = 92.52(69)$ MeV.  In the QGP, the relative abundances of photons, gluons, and light quarks $u, \bar{u},d,\bar{d}$ are in the ratios 1:8:$\frac{9}{4}$, and $s$ quarks have a slightly lower abundance.  These flavor ratios apply both to the thermal $q \bar{q}$ quarks and the ``baryon excess" quarks.   The Baryon Asymmetry of the Universe today, $\eta_0 \equiv  n_b/n_\gamma = (5.8-6.5)\times10^{-10} $ (95\%CL)~\cite{pdg18} amounts to a roughly part-per-billion difference between the $q$ and $\bar{q}$ abundance for each light flavor; to excellent approximation the chemical potential can be ignored for calculating abundances above 100 MeV.   Below the hadronization transition, the most abundant particles besides photons and leptons are pions.  Weak interactions maintain flavor chemical equilibrium, and hadronic and EM reactions like $\pi^+ \pi^- \leftrightarrow \gamma \gamma$ keep hadron abundances in thermal equilibrium well into the low temperature phase.  

 \begin{figure}
\includegraphics[width=\linewidth]{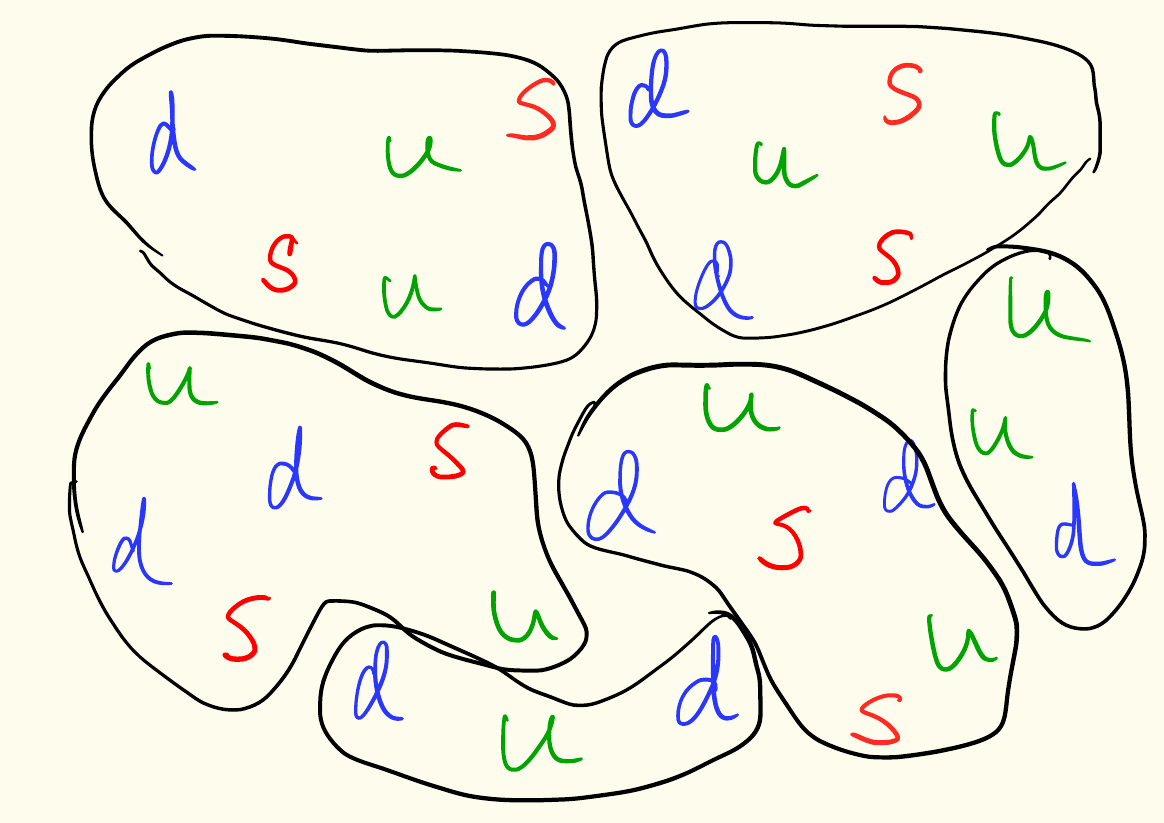}
 \caption{Schematic illustration of how the deficit of $s$ quarks relative to $u,d$ quarks, $\approx 15$\% at the transition temperature, leads to residual baryons.}  \label{fig:endQGP}
\end{figure}

\subsection{Dark Matter abundance: ${\Omega_{DM}}/{\Omega_b}$}
\label{sec:DMtob}
The microphysics of the QGP to hadron transition is not amenable to detailed calculation, but statistical physics determines the relative occupation of states of different energies at any given temperature.  Thus we can estimate the relative abundance of states giving rise to sexaquarks and to anti-sexaquarks, and those giving rise to baryons and anti-baryons, at any given temperature.   There are, in addition, other quark, anti-quark and gluonic states which give rise to mesons but those are not our interest.  We speak below of sexaquarks and baryons, but the story is the same for their anti-particles; 
at these temperatures the baryon-anti-baryon asymmetry is almost negligible. 

In the \sdm\ scenario, DM consists of sexaquarks containing 2 each of $u,d,s$ quarks.   Simply due to their higher mass, the equilibrium fraction of strange quarks and antiquarks, $f_s \equiv (n_s + n_{\bar{s}})/\sum_{i=1}^3(n_i + n_{\bar{i}})$ is lower than that of up and down quarks and antiquarks.  Over the relevant temperature range, 140-170 MeV, the fraction of $s$ quarks in thermal equilibrium varies from 30-31\% with the remaining 70-69\% being equally $u,d$. 

If every $s$ in the quark-gluon plasma were in a sexaquark and baryons were only formed from the left-over $u,d$ quarks, the density of \s's would be $\frac{f_s}{2} \, n_q$, where $f_s$ is the fraction of quarks that are $s$'s and $n_q$ is the total density of quarks.  Since each $s$ quark in an \s\ is accompanied by a $ud$ pair, the density of left-over baryons would be $\frac{(1-3 f_s)}{3} \, n_q$.   As the temperature drops from 170 to 140 MeV, $3 f_s$ changes only slightly, from 0.964 to 0.948.   

Not every strange quark is in a sexaquark, so we introduce $\kappa_s $, the efficiency with which $s$ quarks are trapped in sexaquarks.  Thus we have 
\be
\label{ODMoverOb}
\frac{\Omega_{DM}}{\Omega_b} = \frac{y_b \,  \kappa_s \, 3 f_s}{1 - \kappa_s\, 3  f_s}~,
\ee
where $y_b \equiv m_S/(2 m_p)$ is near 1.

We can estimate $\kappa_s $ as follows.  First consider production of \s's.  Even at the level of 1-gluon exchange, which provides a good qualitative accounting of most hadron mass splittings \cite{DGG75}, there is a strong hyperfine attraction between $uuddss$ quarks in the sexaquark (color-, flavor- and spin-singlet) configuration \cite{MAC,jaffe:H}.  This perturbative attraction is present independently of whether the quarks are in an isolated, zero-temperature \s\ particle, quark nuggets, or are in the QGP.   Thus when the strongly attractive sexaquark configuration of quarks occurs by chance in some spatial region of the QGP, it will be energetically favored and linger in that state.  Quarks in configurations which are not energetically favored will continue their random rearranging.  

Because the chemical potential is negligible, statistical physics tells us that the relative probability of finding two $s$ quarks in an \s-like state compared to finding them in a state consisting of two separate (hyperon-like) 3-quark states, is exp$(\Delta E)/T$ where $\Delta E$ is the energy splitting of the two configurations.   When hadronization occurs, the \s-like color singlet states become \s's and other color singlets become mesons, baryons and anti-baryons;  configurations which are not color singlets continue rearranging and form new color-singlet combinations which then become hadrons.

We can estimate $\Delta E$ and hence $\kappa_s$ using physical masses of nucleons, hyperons and the hypothesized mass of the \s; this approximation gives

\be
\label{kappa}
\kappa_s(m_S,T) = \frac{1}{1 + \left( r_{\Lambda,\Lambda} +  r_{\Lambda,\Sigma} + 2  r_{\Sigma,\Sigma} + 2  r_{N,\Xi}\right)} ~
\ee   
where $r_{1,2} \equiv {\rm exp}[-(m_1 + m_2 - m_S)/T] $ and the coefficients of the different terms are the number of combinations of the given baryon states containing $uuddss$.   The leading uncertainty due to confinement and chiral-symmetry breaking cancels, to the extent that the presence or absence of the quark and gluon condensates shifts the masses of the \s\ and octet baryons together.  

\begin{figure}[tbp]
\centering
\includegraphics[trim={1.5in .6in .7in  0in},clip,width=\linewidth]{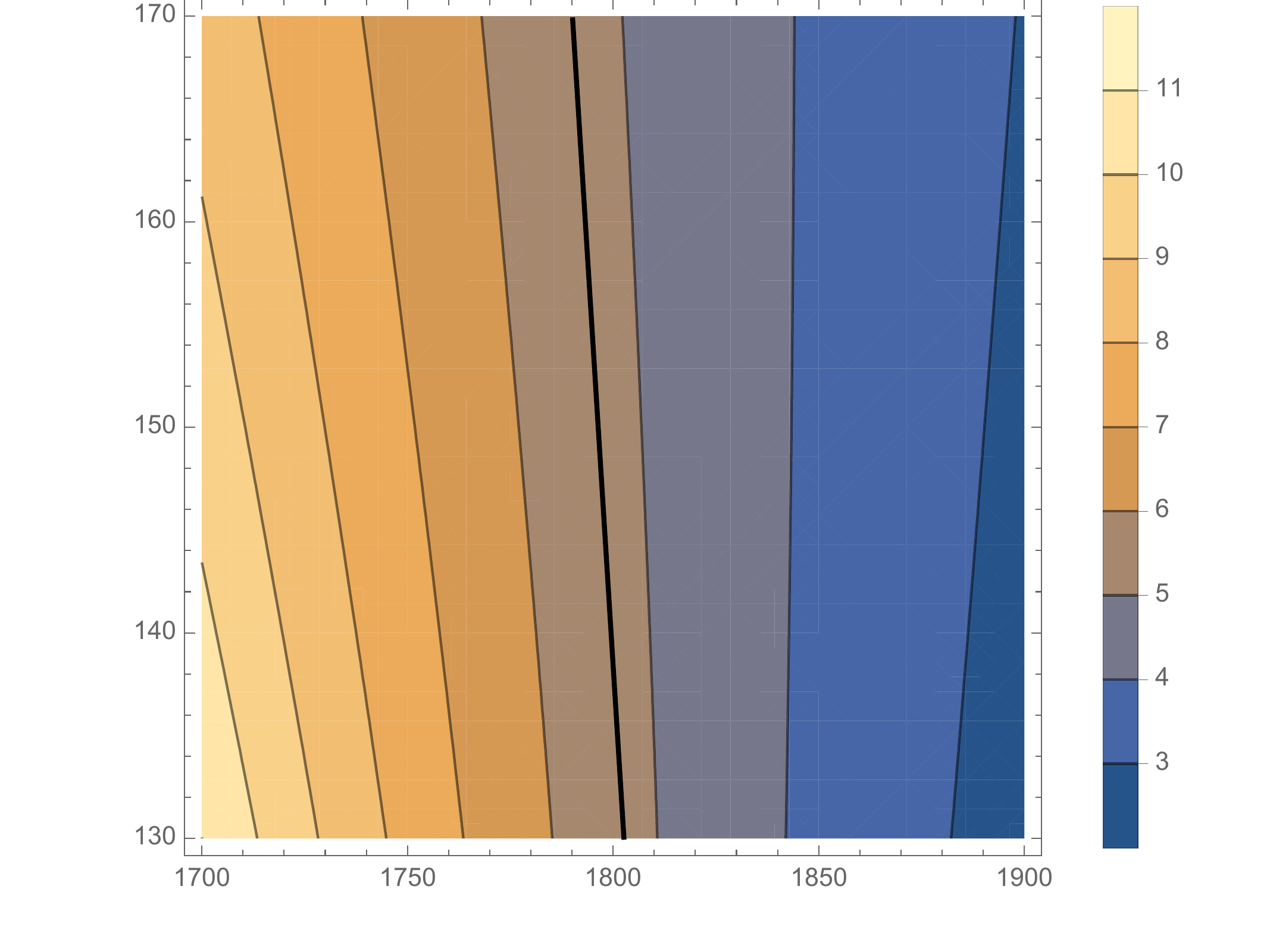}
\vspace{-0.18in}
\caption{$\Omega_{DM}/\Omega_b$ versus $m_S$ (in MeV, vertical axis) and the effective freezeout temperature (in MeV, horizontal axis).   The predicted value ranges from 3 (blue) to 8 (light tan);  the measured value $5.3 \pm 0.1$ is indicated by the black line.  }

		\label{ODMOb}
	\vspace{-0.1in}
\end{figure}

Idealizing the production of DM as occurring at a single effective temperature somewhere in the 140-165 MeV range, and using Eq. (\ref{kappa}) to calculate $\kappa_s(m_S)$, leads to the values of  \ODMOb\ shown in Fig. \ref{ODMOb}.  The predictions are within a factor-2 of the measured ratio $\Omega_{DM}/\Omega_b = 5.3 \pm 0.1$ over the entire plane.   The mild dependence of \ODMOb\ across Fig. \ref{ODMOb} follows from the fact that $f_s$ and $\kappa_s$ have the opposite behavior as $T$ changes, so the product $\kappa_s f_s$ entering Eq. (\ref{ODMoverOb}) varies relatively little, making the prediction robust to uncertainties in $T_{\rm eff}$.   Thus the observed ratio of DM and baryons  is a very robust consequence at the $\mathcal{O}(1)$ level of the sexaquark DM model.
It is noteworthy that an analogous Freezeout Approximation treatment of nucleosynthesis in central Relativistic Heavy Ion Collisions gives an excellent accounting of abundances over 9 decades in branching fraction \cite{Andronic+HIC_Nature18} with $T_{\rm fo} = 156$ MeV.   For reference, the exact observed value $\Omega_{DM}/\Omega_b = 5.3$ is obtained for $T_{\rm eff} = 156$ MeV with $m_S = 1860$ MeV, while for $m_S = 2 m_p$,  $T_{\rm eff} = 150$ MeV.  

One might be tempted to take $T_{\rm fo} = 156$ MeV from relativistic heavy ion collisions and infer $m_S$ from the observed value of  \ODMOb.   However that would not be correct because there are $\mathcal{O}(1)$ uncertainties inherent in the analysis.  Most importantly, the turn-on of confinement and chiral symmetry breaking as the Universe cools is almost static and the medium is almost homogeneous, whereas in a heavy ion collision the plasma expands into the vacuum on a short timescale so that $T_{\rm eff}$ can be somewhat different from $T_{\rm fo}$ inferred from fitting heavy ion data~\cite{Andronic+HIC_Nature18}.   Furthermore, we ignored possible contributions of resonances to particle abundances on account of the long timescale of the Early Universe process, whereas their inclusion improves the fit for Heavy Ion Collisions~\cite{Andronic+HIC_Nature18}.   We also used $T=0$ masses in vacuum to estimate $\kappa_s$ via Eq. (\ref{kappa}), whereas hadron masses in a nuclear medium are known from experiment and lattice QCD~\cite{KoKochLi97} to differ by $\approx 10$\% or more from their $T=0$ values across this temperature range.  
If a sexaquark would be discovered so that $m_S$ is fixed, the 2\% precision with which \ODMOb\ is known will give insight into how QCD condensates and the energy difference of sexaquark-like and hyperon-like states evolve with temperature.  
 
Implicit in the above discussion, is that the value of \ODMOb\ established in the hadronization transition persists to the recombination epoch where it is measured \cite{PlanckCosmo15}.  
For sexaquark DM, non-destruction requires the cross section for reactions such as $\pi S \rightarrow \Sigma \Lambda$, $K S \rightarrow p \Lambda$ and $\Lambda \Lambda \rightarrow S \pi \pi$ to be small.  This is consistent with the transition amplitude $\tilde{g}$ between an \s\ and two baryons being suppressed, as discussed in the next section.

\subsection{Durability of the \s\ in the hadronic phase}
\label{sec:durability}

If breakup processes such as $\pi S \leftrightarrow \Sigma \Lambda$, $K S \leftrightarrow p \Lambda$ and $\Lambda \Lambda \leftrightarrow S \pi \pi$ had a typical hadronic rate, these rates would be fast compared to the Hubble expansion rate at the temperatures of interest, $T \sim 150$ MeV.  In that case, the \s\ would quickly come into chemical equilibrium with baryons and the chemical potentials would satisfy $\mu_S = 2 \mu_b$.   However $\mu_S = 2 \mu_b$  at $T \sim 150$ MeV, along with $m_S \approx 2 m_p$,  implies $\lesssim 10^{-7}$ of the baryon number is carried by \s's, so an initial \sdm\ excess comparable to the baryon excess would quickly disappear.    

Thus we need to estimate the breakup rate of \s's.  We use lowest order meson-baryon effective field theory extended to include the sexaquark and $SBB'$ vertices, and elaborate the discussion in \cite{fudsDM18}.  The Lagrangian of the low energy effective field theory describing the interactions between the flavor-singlet \s\ and flavor-octet baryons can be written
\be
\mathcal{L} = \frac{\tilde{g}}{\sqrt{40}} \overline{\psi_B} \gamma_5 \psi_{B'}^c S + g_{S SV} S^\dagger \partial_\mu S V^\mu + h.c.
\label{eq:L}
\ee
where $V^\mu$ is shorthand for the flavor-singlet linear combination of $\omega, \phi$ vector meson fields.  The second term governs \sdm\ scattering cross sections and we will return to it in later sections.  \s\ breakup to baryons is governed by the first term.  In Sec. \ref{sec:bu} we will estimate $\tilde{g}$ theoretically and constrain it from observations.  Here, we determine the maximum value of $\tilde{g}$ consistent with \sdm\ surviving in the hot Early Universe.  

The color-flavor-spin wavefunction of the totally antisymmetric 6-quark color-flavor-spin singlet state is derived in~\cite{fwSwfn20}, where it is shown that the quarks in an \s\ have highly entangled wavefunctions such that only a small fraction (1/5) can be decomposed into color singlet pairs of $B=1$ states.  Projecting the \s\ wavefunction in terms of quarks onto that of a pair of color-singlet octet baryons, gives the color-flavor-spin wavefunction-overlap
\begin{align}
\label{SBB'}
<\, S \, | \,\Lambda \Lambda> & = \,<\, S \, | \,\Sigma^0 \Sigma^0> \, = - <\, S \, | \,\Sigma^+ \Sigma^->  \\
					& =  - <\, S \, | \, n \, \Xi^0  > \, = \, <\, S \, | \, p \, \Xi^-  > = \frac{1}{\sqrt{40}}, \nonumber
\end{align}
motivating the $1/\sqrt{40}$ in the denominator of Eq. (\ref{eq:L}) so that $\tilde{g}$ is the dynamical transition amplitude $ <B \, B'| \mathcal{H}_{\rm QCD}\,| S>$ between quarks in the \s\ and those in spatially separated 3-quark states.

The breakup processes with the highest rates are $\pi^\pm S \rightarrow \Sigma^\pm \Lambda$ and $K^+ S \rightarrow p \Lambda$ with amplitudes 
\be
\label{ampLamSig}
\mathcal{M}_{\pi^\pm S \rightarrow \Sigma^\pm \Lambda} \approx \frac{f \tilde{g}}{\sqrt{120}} (1-\alpha) m_\Lambda m_\Sigma v_{\rm rel} \left(\frac{1}{m_\Lambda^2} - \frac{1}{m_\Sigma^2} \right) ;
\ee
\be
\mathcal{M}_{K^+ S \rightarrow p \Lambda} \approx \frac{f \tilde{g}}{\sqrt{120}} m_\Lambda m_p v_{\rm rel} \left( - \frac{(1+ 2 \alpha)}{m_\Lambda^2}  -  \frac{(4 \alpha - 1)}{m_\Xi^2} \right)  .\nonumber
\ee
Here $f = 0.952$ and $\alpha = 0.365$ are parameters characterizing the meson-baryon couplings, taken from \cite{stoksMBcouplings97} where they are fit to data, and  $v_{\rm rel}$ is the relative velocity in the final state.   The $v_{\rm rel}$ factor arises because the baryons must have $L=1$ in order to satisfy parity and angular momentum conservation and Fermi statistics, given that the $\pi/K$ is a pseudoscalar, the \s\ is an even parity, spin-0 particle, and the intrinsic parity of a pair of baryons is +1.  

Performing the thermal average \cite{cannoni17} to determine the \sdm\ breakup rate shows that  $\Gamma(K^+ S \rightarrow p \Lambda) = n_{K^+}(T) <\sigma_{K^+ S \rightarrow p \Lambda} \, v>$ is about two orders of magnitude larger than $\Gamma(\pi^\pm S \rightarrow \Sigma^\pm \Lambda)$, and $\Gamma(\Lambda \Lambda \rightarrow S \pi \pi)$ considered in \cite{kt18}.  The suppression of $\pi^\pm S \rightarrow \Sigma^\pm \Lambda$ results from the cancelation between the contributions of virtual $\Lambda$ and $\Sigma$ in (\ref{ampLamSig}) due to the opposite sign of $<\, S \, | \,\Lambda \Lambda>$ relative to $<\, S \, | \,\Sigma^+ \Sigma^->$ in Eq. (\ref{SBB'}), while $\Lambda \Lambda \rightarrow S \pi \pi$ is suppressed by 3-body phase space.

The Hubble expansion rate is greater than the dominant breakup rate $\Gamma(K^+ S \rightarrow p \Lambda) = n_{K^+}(150 {\rm MeV}) <\sigma v>$, for $\tilde{g}^2 <  4 \times 10^{-12} $.  In the next section we discuss theoretical expectations and experimental constraints on $\tilde{g}$.  As will be seen, the condition  $\tilde{g} \lesssim 2 \times 10^{-6}$  is well within the expected range.  The assertion of \cite{kt18} that dark matter cannot be dibaryonic, based on assuming a conventional hadronic breakup rate, is therefore not correct.  The related but earlier work of \cite{strumia+18} also assumes the relic \sdm\ abundance is determined by the conventional thermal freezeout mechanism, leading them to conclude that a dibaryon mass of order 1.2 GeV is required for dibaryonic dark matter.  Instead, the mechanism pointed out in \cite{fudsDM18} and reviewed above, makes essential use of the fact that chemical equilibrium is not assured in the hadronic phase.

\section{\label{sec:bu} Sexaquark breakup interactions}

\subsection{\label{sec:buTh} Modeling \gsbb\ }

The effective hadronic Yukawa coupling for sexaquark breakup, \gsbb, is
\be
\label{gsbb}
\tilde{g}\, \bar{u}_B  \gamma_5 v_{B'} \equiv <B \, B'| \mathcal{H}_{\rm QCD}\,| S>~.
\ee
We work in the approximation that this transition amplitude is independent of the baryon masses.  Actual transitions of physical interest such as \s\ decay require additional factors of $\mathcal{H}_{\rm w}$, but these weak interactions are perturbative so to good approximation they factorize from the QCD part of the transition.  

 \begin{figure}

\includegraphics[
width=\linewidth]{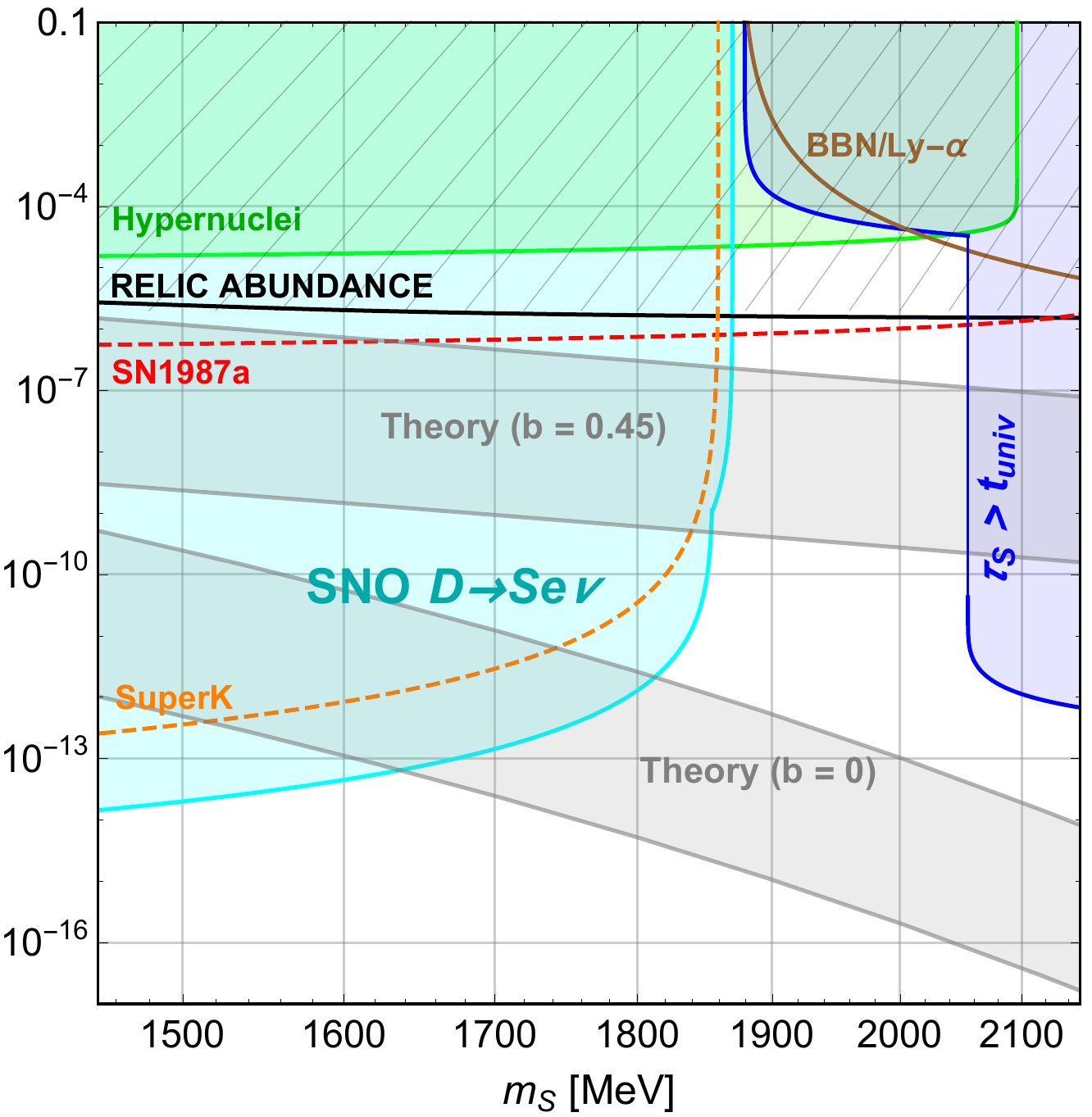}

 \caption{Predicted and excluded regions for \gsbb\ as a function of $m_S$;  the horizontal black line is the maximum value of \gsbb\ compatible with non-destruction of sexaquark DM in the hot hadronic phase (Sec. \ref{sec:durability}).  The grey bands show the range of theory predictions for \gsbb, for the extreme values $b$ = 0 and 0.45 (Eq.~\ref{rSb}), using the central hard core radius from fits to scattering data, $r_c = 0.4$ fm (Sec. \ref{sec:buTh}).  The theory predictions shift up (down) by a factor $\approx 10^2$ ($10^{-3}$) for $r_c = 0.3\,(0.5)$ fm.    The width of the theory bands reflects the uncertainty range for the tunneling suppression.  The green shaded region is excluded by non-production of \s\ in hypernuclear experiments~\cite{fzNuc03}.  The cyan shaded region is excluded by our analysis of the stability of deuterons using SNO, reported in Sec. \ref{sec:buSNO}, taking the  Bethe-Salpeter momentum scale $P=100\,$MeV; the cyan \gsbb\ limit scales as $(100\, {\rm MeV} /P) ^3$.  The dashed orange line is the upper limit from an estimated SuperK background rate \cite{fzNuc03}.  
The region above the blue line is excluded by requiring the \s\ lifetime to be greater than the age of the Universe, in the mass range where decay $S\rightarrow n n$ is kinematically allowed.  The dashed blue line is a stronger limit that could be obtained from our SNO limits on $S \rightarrow nn,\,n\Lambda$ (Sec. \ref{sec:buUnstab}) if the local number density of free \sdm\ particles were $10^{14}{\rm cm}^{-3}$ as found in \cite{nfm18}, however such a high local number density of free \s's can now be excluded by the discussion in Sec. \ref{sec:DDconstraints}.  The brown line comes from requiring that deuterium observed in damped Ly-alpha clouds be consistent with the BBN prediction plus a component from $S\rightarrow D e^- \bar{\nu}$, within 3-sigma; the constraint is weak because this decay is suppressed by 3-body phase space. The red line is the limit based on SN1987a cooling following  the analysis of \cite{mcdermott+SNe18}; it is dashed because this use of SN1987a has been called into question by \cite{blum+SN1987a_19}. In general, limits dependent on some additional, possibly invalid assumption are shown with dashed lines.  
}  \label{fig:gsbb}
\end{figure}

 \begin{figure}
 \includegraphics[
width=\linewidth]{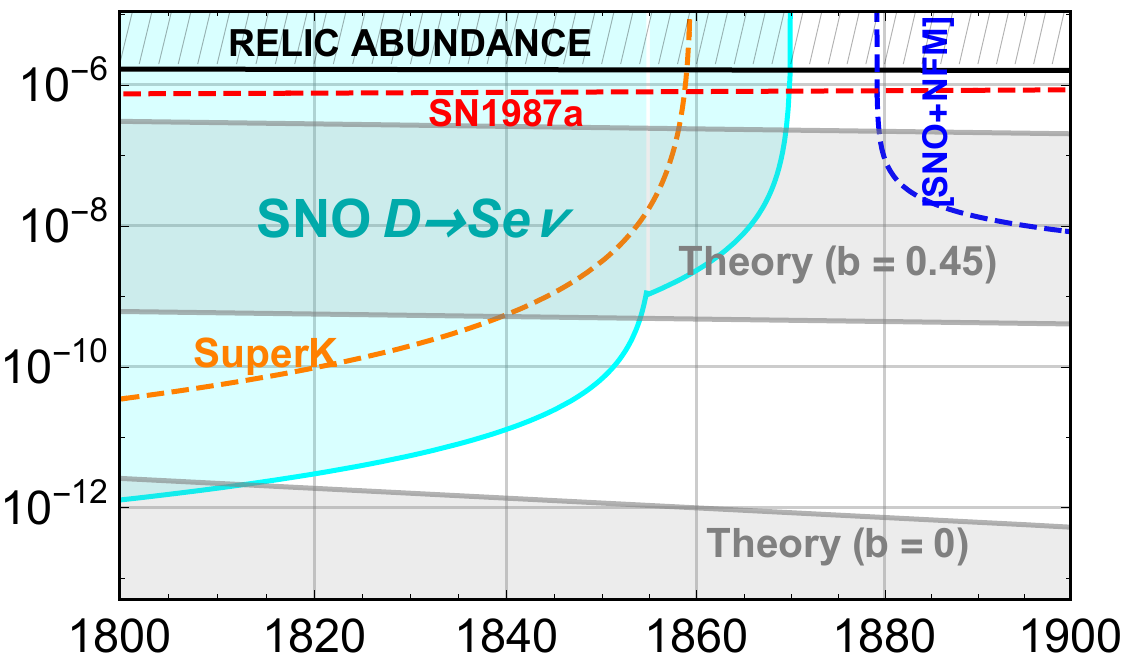}
 \caption{A blow-up of the 1800-1900 mass range in previous figure. }  \label{fig:gsbbBU}
\end{figure}

The QCD transition amplitude between \s\ and two baryons, distilled into the effective field theory parameter \gsbb\ in Eq. (\ref{gsbb}), describes the process by which each quark moves from an initial position in the \s\ into a final position in one of the baryons, in the field of the other moving quarks, integrated over all possible paths.  The hard-core repulsion of baryons at short distances, responsible for the relative incompressibility of nuclear matter, implies a high potential barrier to the transition between the initial separated configuration and final compact configuration.   Furthermore, if the spatial extent of the \s\ wavefunction is of order or smaller than the effective radius of the hard core repulsive B-B potential, the overlap of the initial and final spatial wavefunctions is small as well.  These are distinct effects, as can be appreciated by recognizing that even if the \s\ were large compared to the hard core radius and had a similar spatial extent to a deuteron (2 fm), the transition can be highly suppressed if the barrier to rearrangement is high, as exemplified by crystalline phase transitions and protein folding.

The evaluation of \gsbb\ entails relativistic, non-perturbative QCD dynamics whose modeling from first principles is far beyond current theoretical reach.  When the number of quarks is larger than 3, even calculating static properties like masses and magnetic moments is challenging, even if the quarks are effectively non-relativistic, which is not the present case.  
Recognizing the large uncertainties, we proceed to estimate \gsbb\ as a product of the wave-function overlap and a tunneling suppression factor.  

The procedure for calculating the wavefunction overlap was developed in \cite{fzNuc03} where several different nuclear wavefunctions, hard-core radii and model parameters for quark distributions in the Isgur-Karl wavefunction were explored.  The value of the most significant parameter -- the hard-core radius in the nucleon-nucleon potential,  $r_c$ --  is uncertain, in part because probing it takes relatively large momentum transfer where a simple potential model description begins to fail.  The Hamada-Johnston potential has a hard core of 0.343 fm \cite{HamadaJohnston62} and the Reid hard core is $r_c = 0.429$ fm \cite{Reid68}; with modern data the hard core radius could possibly be stretched to 0.5 fm, but most likely not more (R. Wiringa, private communication).   

The tunneling suppression, $e^{-\mathcal{S}}$, can be estimated as follows.  In natural units, the action for a single quark $\mathcal{S}_q \sim \Delta E \, t$ with $t \sim $ fm, the time to cross the system (in natural units), and $\Delta E = 100-300$ MeV, the QCD scale.  Summing over the 6 quarks, the total action for the transition can then be estimated as $\mathcal{S} =  3-9 $, for a tunneling suppression of 0.05 -10$^{-4}$.  This may underestimate the tunneling suppression because lattice gauge calculations show the inter-baryon repulsive potential grows rapidly at short distance, reaching 300 MeV at the shortest distance reported in \cite{HALQCDS-2}, $r \approx 0.3$ fm, so the typical action per quark may actually be greater than 1.5.   

The grey bands in Fig. \ref{fig:gsbb} show \gsbb\ calculated as the product of the tunneling suppression and the spatial overlap, using the consensus value $r_c = 0.4$ fm.  The band shifts up (down) by a factor $10^2$ ($10^{-3}$) for $r_c = 0.3\,(0.5)$ fm.   In the lower band, we have taken the radius of the \s, which enters the calculation through the Isgur-Karl spatial wavefunction, to be its Compton wavelength.  This is motivated because the \s\ does not couple to pions or other light particles and therefore is not spread out by a pion cloud, unlike baryons which couple to pions having Compton wavelength $> 1$ fm.  To show the extreme alternative, in which the \s\ is as strongly coupled to the mediator as the nucleon is to the pion, the upper grey band is calculated with $b=0.45$. 
In principle, determination of the Yukawa parameter \asn\ discussed in the next section would enable the range of $b$ to be narrowed.
 For further details and plots showing sensitivity to secondary parameters, see \cite{fzNuc03,wfS20}.  
 
 We note that two calculations in the literature did not take proper account of the hard core radius,  effectively causing them to overestimate \gsbb.  Ref. \cite{strumia+18} follows the analysis of \cite{fzNuc03} but uses a wavefunction fit to large distance data which does not incorporate hard-core-sensitive information, thus the overlap derived in \cite{strumia+18} is much larger and not realistic.  Ref. \cite{mcdermott+SNe18} ignores the short distance repulsion altogether and hence greatly overestimates the impact of sexaquarks on the cooling rate of SN1987a.

Although the estimated range for \gsbb\ presented here has large uncertainties, it is far below the $\approx 2 \times 10^{-6}$ value shown by the black line in Fig. \ref{fig:gsbb}, where sexaquark breakup would decrease the DM to baryon ratio established in the QGP-hadronization transition.   Therefore the prediction \ODMOb\ $\approx 5$ of the previous section is a robust prediction of \sdm, and its agreement with the observed \ODMOb$= 5.3 \pm 0.1$ is a significant point in favor of the \sdm\ scenario.

In the next subsections we consider direct experimental constraints on \gsbb.

\subsection{\label{sec:buSNO} Experimental constraints on nuclear decay into sexaquark}
For $m_D - m_e  < m_S <  m_D + m_e$, the \s\ and all nuclei are absolutely stable.  In this range, the only constraint on the sexaquark breakup or formation vertex $\tilde{g}$ comes from the observed lifetimes of double-Lambda hypernuclei \cite{hyper2011,hyper2012,fzNuc03,fS17}.  The excluded region from hypernuclei is shown as the green region in Fig. \ref{fig:gsbb}. 

If $m_S  < m_D - m_e = 1876.12$ MeV, deuterium can decay via $D \rightarrow S e^+ \nu_e$.  The quark-level decay diagrams contain two $u \rightarrow s + W^+$ vertices, with one $W^+$ emitted and another being absorbed, $W^+ \, d \rightarrow u$, for a net quark level transition $ u d \rightarrow s s W^+$ with amplitude $\sim G_F^2 {\rm sin}^2 \theta_C \, {\rm cos} \theta_C$.  
Since we cannot perform a detailed quark-level analysis we capture the essential features of the 3-body phase space by taking the energy dependence to be that of neutron beta decay; inserting the factors from the amplitude and setting ${\rm cos} \theta_C$ to 1: 
\begin{equation}
\label{dGamdE}
    \frac{d\Gamma}{dE_e } = \frac{\tilde{g}^2 \, G_F^4 \, {\rm sin}^4 \theta_C \, 
    P^6 }{120\pi^3 m_\textrm{D} m_S} \sqrt{E_e^2-m_e^2} E_e(m_\textrm{D}-m_S-E_e)^2.
\end{equation}
The $P^6$ factor would emerge from an integration of the quark amplitude over the Bethe-Salpeter wavefunctions of nucleons and \s, if those were known and included in the treatment.  For our numerical estimates we take $P = 100$ MeV unless otherwise stated, characteristic of the QCD scale.  Clearly this is a major source of uncertainty given the high power  of $P$ involved. 
The total decay rate is obtained by integrating Eq. (\ref{dGamdE}) from $m_e$ to $m_D - m_S$.

The SNO detector contains 1000 tons of heavy water.  The positrons produced in $D\to Se\nu$ would be detected via their Cherenkov light if the positron's energy is above the SNO $5.5$ MeV threshold~\cite{SNO}.  (SNO was built to detect electrons produced by solar neutrinos, $\nu_e+D \to2p+e^-$, but positrons are functionally equivalent to electrons because the spectrum of Cherenkov radiation is only sensitive to charge-squared.)  In 391.432 days, SNO has observed $N_{\rm obs} \approx$ 2465 $e^\pm$ events with kinetic energies in the 5-20 MeV range, and none above.  Let $f(E_{\rm th})$ be the fraction of the spectrum (Eq. \ref{dGamdE}) above some specified electron total energy threshold $E_{\rm th}$.  We obtain limits on $\Gamma$ and \gsbb\ by requiring (suppressing the dependence of $\gamma$ and $f$ on $m_S$): 
\begin{equation}
\begin{cases}
   (f(5.5)-f(20))\times N_0(1-e^{-\Gamma t}) < N_{\textrm{obs}}, \\
    f(20)\times N_0(1-e^{-\Gamma t}) < 2.44,
\end{cases}
\end{equation}
whichever is stronger.   The 2.44 in the second equation is the 90\% CL upper limit when no events are seen; given the large value of $N_{\textrm{obs}}$ and uncertainties in the analysis, we do not correspondingly adjust $N_{\textrm{obs}}$.

The analysis outlined above using SNO data and taking $P = 100$ MeV, gives the cyan exclusion region in Fig. \ref{fig:gsbb}.  As a result of the uncertainties in calculating the weak-decay amplitude in terms of \gsbb, the boundary curve should be regarded as a best-estimate indication of the limit on $\tilde{g} (P /{\rm 100 \, MeV} )^3$.  Not surprisingly, the limit on $\tilde{g}$ becomes dramatically stronger as $m_S$ drops and phase space for $D$ decay opens up.  Due to the wide range of uncertainty in the theory predictions, inverting the bound from $D$ lifetime to get a lower bound on $m_S$ is not very meaningful  -- especially keeping in mind that $r_c > 0.4\,$fm is not excluded.  Nonetheless, one is tempted to deduce provisionally at least, that $m_S < 1800\,$MeV is disfavored even without invoking theoretical prejudices against low masses.


\subsection{\label{sec:buUnstab}  Constraints on \gsbb\ from sexaquark lifetime}

If $m_S > m_D - m_e  $, the sexaquark is not absolutely stable and can decay via $S \rightarrow D e^- \bar{\nu}$. This process provides an additional astrophysical source of deuterium and is more significant at low-redshift as more \sdm\ would have decayed. The primordial deuterium abundance $\textrm{D/H}\times 10^5$ is measured in damped Lyman-$alpha$ (DLA) systems at $2.53\pm0.04$~\cite{}, while the BBN theory predicts $2.45\pm0.1$~\cite{}. Thus, the ``excess'' D/H in DLAs within $3\sigma$ is $0.08 + 3\times 0.11=0.41$ where $0.11=\sqrt{0.04^2+0.1^2}$ follows from the propagation of uncertainties. This constrains the fraction of \sdm\ that had decayed til $z\sim2$ or $t\sim 2.6$ Gyr, and the lifetime of $S \to D e \nu$ is $\tau \gtrsim 1.6\times 10^{15}$ yr. Using the the three-body decay rate which can be obtained from Eqn.~(\ref{dGamdE}) with $m_D$ and $m_S$ replacing each other, we can impose upper limits on \gsbb\. The result is shown as the brown curve in Fig.~\ref{fig:gsbb}.

Since three-body decay is strongly suppressed near threshold, a more powerful constraint for most of the mass range comes from $S \rightarrow n n$, or $S \rightarrow \Lambda n$ when that is kinematically allowed.  The analysis is straightforward, proceeding along the lines in the previous subsection but simpler due to the 2-body phase space.  The blue line in Fig.  \ref{fig:gsbb} comes from the constraint that \s\ is longer-lived than the age of the universe.  A stronger limit may be possible, since if $\tau_{\rm DM} = \tau_{\rm Univ}$, 37\% of the DM at recombination would have converted to baryons by $z=0$ with potentially observable implications, but there are other stronger constraints in much of this mass range anyway.  (Decaying DM in scenarios with stronger effects than here have been discussed by~\cite{essigetal13,SlatyerWu16}, but in these models most or all of the DM rest mass is converted to $e^\pm$ or $\gamma$ whereas here only $\mathcal{O}(\Delta M/m_S \approx 0.01)$ is released into EM radiation.)  For $m_S > m_n + m_\Lambda$ about 8\% of the decaying DM mass is converted to radiation (the ultimate $\gamma$, $e$ and $\nu$ decay products from $\Lambda$ decay), also having potentially interesting cosmological effects that we do not pursue here.

The cyan dashed line is the stronger limit which would follow from SNOlab limits on neutron production, if the ambient density of \s's in the SNOlab detector were $n_S = 10^{14} {\rm cm}^{-3}$.  This scenario is motivated by the Neufeld et al. \cite{nfm18} calculation of the DM atmosphere of Earth as a function of DM mass and interaction cross section, in the approximation that the Earth's atmosphere and geophysics have been constant over its lifetime.   With this caveat, the predicted abundance of DM near the Earth's surface reaches $10^{14} {\rm cm}^{-3}$ for DM mass in the sexaquark range.  If such a DM atmosphere consists of free sexaquarks able to decay, the resulting limits on the \s\ decay lifetime are shown by the cyan-dashed line in Fig.  \ref{fig:gsbb}. 

The bound is calculated as follows.  Phase III of the SNO experiment was equipped with an array of ${}^3$He neutron counters~\cite{SNO} and therefore, neutrons produced from $S\to nn$ could be detected. We require
\begin{equation}
    n_S V \times (1-e^{-\Gamma t}) \times 2 \times \epsilon < N_{\textrm{obs}},
    \label{eqn:SnnLim}
\end{equation}
where $V=904.78$ m${}^3$ is the volume of the tank, $\epsilon=0.182$ is the detector efficiency, and $N_{\textrm{obs}}\approx 7000$ is the observed number of neutron events during $t=385.17$ days.   Taking $n_S\sim 10^{14}\, {\rm cm}^{-3}$ to be the ambient density of \sdm\ in SNO, Eqn.~(\ref{eqn:SnnLim}) implies $\tau\gtrsim 4.96 \times 10^{18}$ yr at $90\%$ confidence level.  

However this limit on \gsbb\ is evaded or weakened if the ambient DM is hybridized with nuclei in the Earth and kinematically unable to decay (Sec. \ref{sec:hyb}), if the DM-nucleon interaction is repulsive so dewar exclusions exclude it (Sec. \ref{sec:dewar}), or if the DM atmosphere calculated in Ref. \cite{nfm18} is overestimated, e.g., due to significant disturbances in the temperature profile of the upper atmosphere producing periods of increased for DM evaporation.  

\subsection{Cooling of SN1987a}
\label{sec:SNecooling}
Ref. \cite{mcdermott+SNe18} argued against a deeply bound dibaryon such as the sexaquark, on the grounds that the reaction $\Lambda \Lambda \rightarrow S \, \gamma$, would cause  SN1987a to cool too fast to account for the neutrinos arriving over 10$\,$s.  The general use of this cooling argument has been called into question in Ref.~\cite{blum+SN1987a_19}, proposing another mechanism for producing last neutrinos that is independent of the proto-NS cooling.  Nonetheless, it is interesting to determine what bounds on \gsbb\ could be derived under the assumption of a 10s cooling time.  We follow the analysis of \cite{mcdermott+SNe18} but invert it to find a limit on \gsbb; this bound is shown as the red dashed line in Fig. \ref{fig:gsbb}.  The limit on \gsbb\ needed to satisfy a 10s cooling time is clearly compatible with the predicted range of \gsbb.  The discrepancy between this result and the conclusion of  \cite{mcdermott+SNe18} that an \s\ in this mass range is ruled out by SN1987a, is due to their neglecting the hard-core repulsion between baryons that strongly inhibits production of \s's.

\section{Sexaquark scattering interactions}
\label{sec:DMints}


 \begin{figure}
 \includegraphics[width=\linewidth]{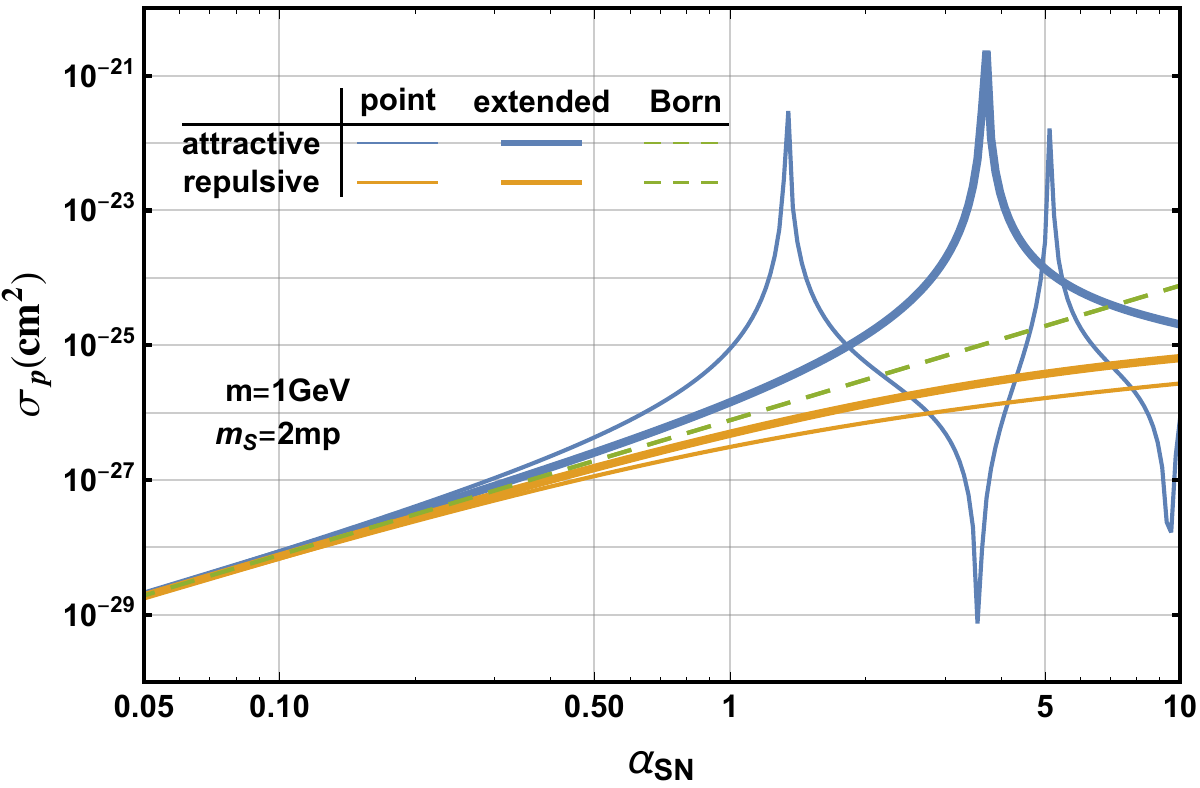}
 \caption{The \s-proton cross section $\sigma_1$ as a function of \asn.  Born approximation becomes inaccurate at the 30\% level for $\alpha > 0.25$; for nuclei Born approximation fails at still smaller $\alpha$.  }  \label{fig:sig_alpha}
\end{figure}

The low-energy interaction between sexaquarks and baryons is mediated by exchange of massive QCD mesonic states, which are constrained by low-energy nucleon and nuclear interactions.  Thus the interactions of \sdm\ with baryons are not arbitrarily adjustable as is often possible in beyond-the-standard model DM scenarios.   In the non-relativistic limit, which is applicable for all the processes we are considering, the elastic scattering of DM by baryons can be described by a Yukawa potential
\begin{equation}
\label{eq:yukawa}
V(r)=-\frac{\alpha}{r}e^{-m  r}~~,
\end{equation}
where the minus sign is for convenience so that $\alpha>0$ corresponds to an attractive force and $\alpha<0$ repulsive.  The attractive case has a richer phenomenology, as discussed below.  Because the discussion here is more general than for \s's, we designate the mediator as $m$ and drop the subscript from \asn\ when not needed for clarity.  

As already noted, since the \s\ is a flavor-singlet, the lightest meson giving a significant contribution is the flavor-singlet combination of $\omega-\phi$ mesons designated $V$.  (The scalar $f_0$, an extended di-meson resonance, is expected be very poorly coupled to the compact \s.)  Taking the mixing angle from \cite{NSC89} we have
\be
|\,V>\, =  0.8\, |\, \omega > -\, 0.6\, |\,\phi> ~,
\ee
 where $m_\omega = 782$ MeV and $m_\phi = 1020$ MeV \cite{pdg18}.    The coupling strength \asn\ 
 \be
 \alpha_{SN} \equiv {g_{SSV} \, g_{NNV}}/{4 \pi}~ 
 \ee
may be as large as $\mathcal{O}(1)$, as is typical for strong interaction processes, or it may be much smaller depending on \gssv, the coupling of $V$ to \s\ appearing in the low energy effective Lagrangian, Eq. (\ref{eq:L}). 

 From \cite{NSC89}, $g_{NNV}/\sqrt{4 \pi} = 2.5$, but we stress that modern analyses of low energy baryon-baryon scattering such as \cite{NSC89,ESC19} are much more sophisticated than a simple one-meson-exchange treatment and include many exchanges and other effects.  Thus while the parameters in those analyses have small error bars, used out of the context of those full analyses a parameter such as $g_{NNV}/\sqrt{4 \pi} = 2.5$ should be considered to have $\mathcal{O}(1)$ uncertainties and just considered as a guide for our problem.   An estimate of the minimum coupling strength range of $g_{SSV}$ might be to rescale $g_{NNV}/\sqrt{4 \pi} = 2.5$ by the square root of the ratio of the size of the \s\ and $V$ (taking the $N$ and $V$ to be fully overlapping and strongly coupled).  With $r_S$ as small as 0.1 fm, and $r_V \approx 1$ fm, this would suggest $g_{SSV}$ a factor 30 smaller than  $g_{NNV}$.  Adding a margin of uncertainty in both directions, we focus our attention on the domain $0.001 < \alpha_{SN} < 10$.   

An exact analytic solution for the Yukawa potential scattering problem does not exist and Born approximation does not apply for the parameter space we are interested in so that a full numerical solution is necessary.   Furthermore, since nuclei are extended, we need to solve the Schroedinger problem for the extended potential obtained by smearing the Yukawa over a hard sphere of radius $R_A \approx R_0\, A^{1/3}$ fm.  We took $R_0 =1.0 $ for the calculations presented in this paper, but models in the literature have values up to 1.25.  Sensitivity of our results to the value of $R_0$ will be reported in ~\cite{xf20}, where details of our calculational methods are given.  We rely heavily on techniques from \cite{TulinYuZurek13}.  Note that for a given model of the nuclear wave function the constraints on \asn\ derived in this and following sections can generally be determined to higher precision than 1 significant figure, but on account of our simplistic approximation that the nucleus is a hard sphere of radius $A^{1/3}\,$fm, reporting higher precision would be misleading.

\begin{figure}
 \includegraphics[width=\linewidth]{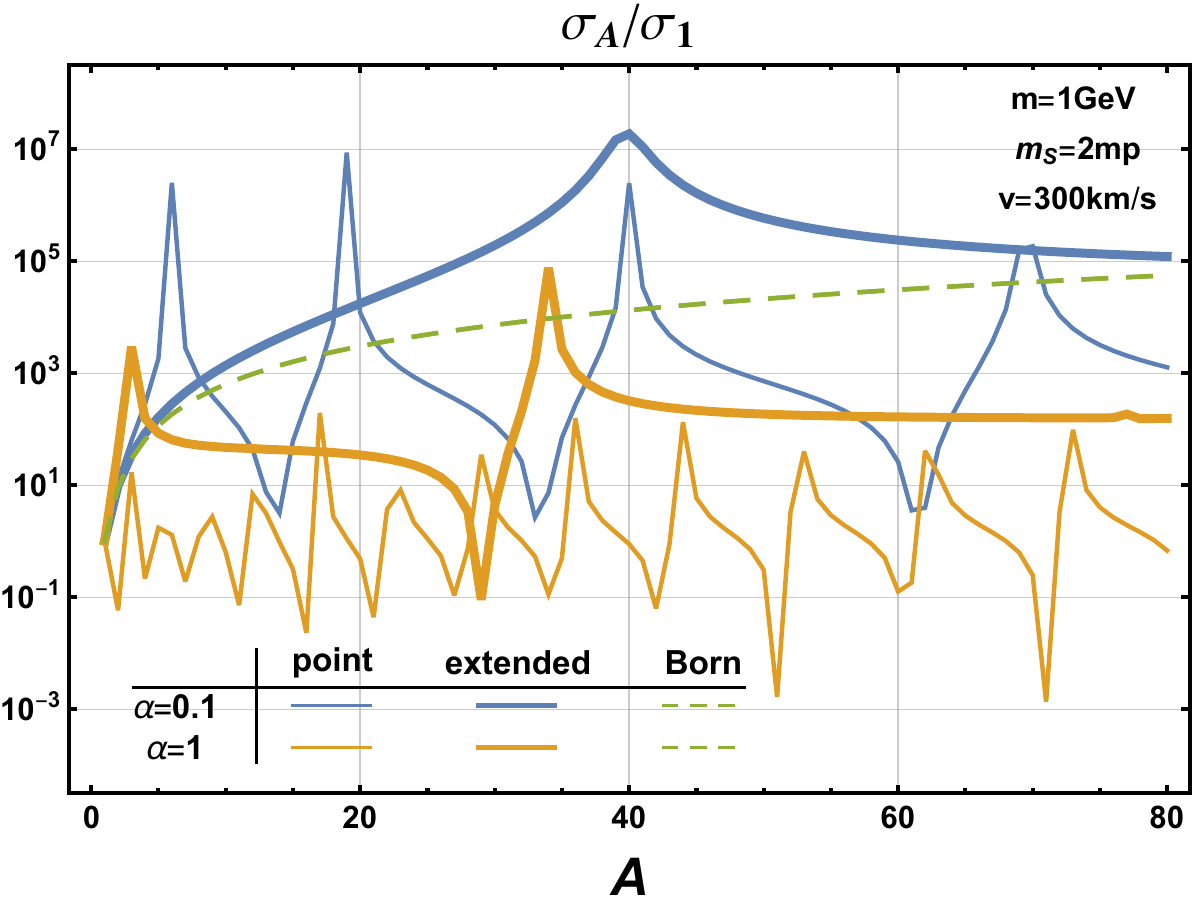}
 \caption{$\sigma_{A}/\sigma_{1}$ versus $A$ for $v = 300$ km/s and $ \alpha_{SN} = 0.1$ and 1.0 (blue, yellow), for an attractive Yukawa interaction and point and extended nuclei (thin and thick curves, respectively); the Born approximation ratio is shown in green-dashed.   Born approximation fails badly for all cases.
}  \label{fig:Ascaling}
\end{figure}

 \begin{figure}
 \includegraphics[width=\linewidth]{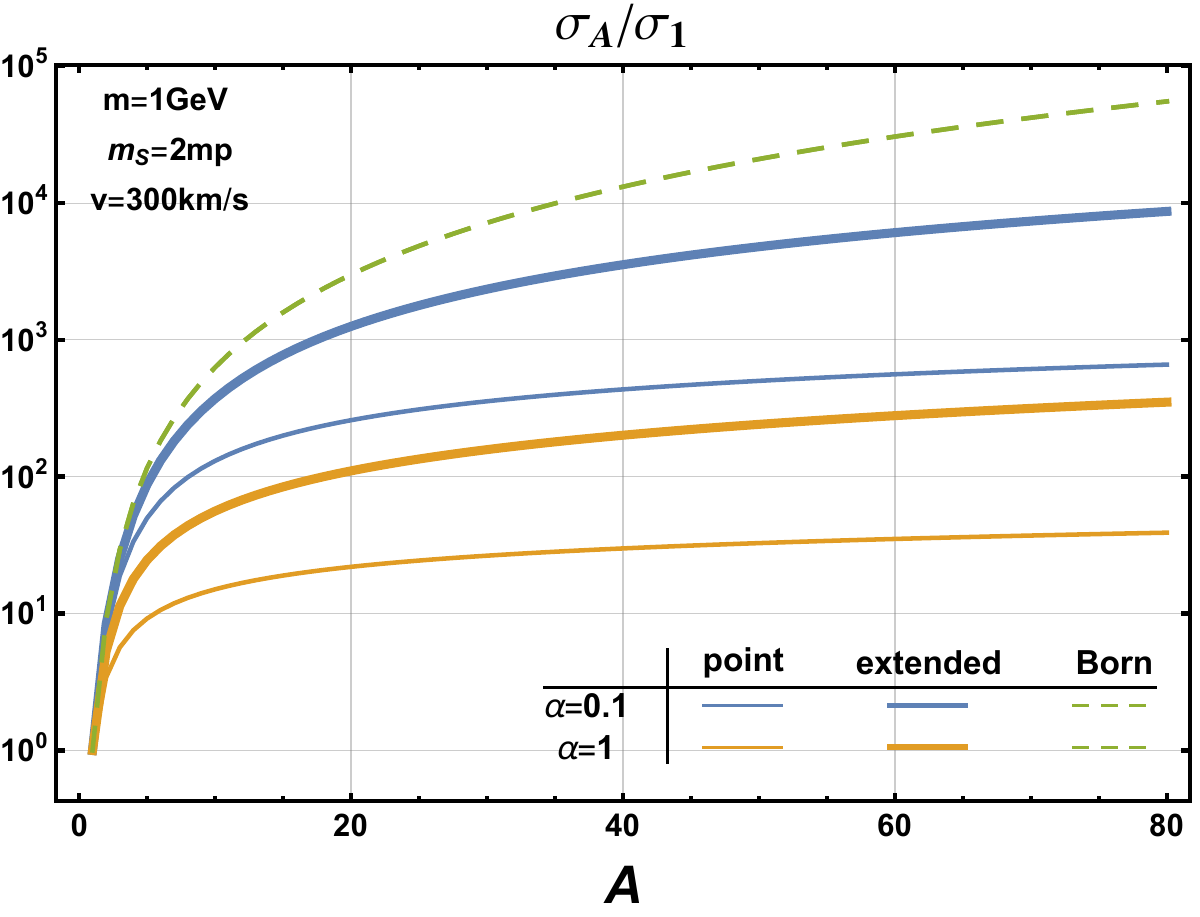}
 \caption{As in Fig. \ref{fig:Ascaling} but for a repulsive interaction.  Born approximation overestimates $\sigma_{A}/\sigma_{1}$ in all cases, by a factor $\approx 40$ for Si with \asn\ = 1 and realistic nuclear spatial distribution.
}  \label{fig:Ascaling_repulsive}
\end{figure}

Figures ~\ref{fig:sig_alpha} ,~\ref{fig:Ascaling} and \ref{fig:Ascaling_repulsive} show the first crucial result:  the inapplicability of the Born approximation over a large portion of parameter space.  Fig.  ~\ref{fig:sig_alpha} shows how the DM-proton cross section, $\sigma_1$, depends on $\alpha$ for extended and point sources and attractive and repulsive interactions.  (We abbreviate $\sigma_{SA} \rightarrow \sigma_A$.)  For $\alpha \gtrsim 0.1$ the Born approximation is inaccurate, with the true cross section being lower than Born approximation for the repulsive case.  For an attractive interaction the true cross section is up to a factor $10^5$ higher and $10^4$ lower than Born, as $\alpha$ ranges from 1 to 5 in the pure Yukawa case; for an extended source such extreme deviations are pushed to larger $\alpha$.

Figures~\ref{fig:Ascaling} and \ref{fig:Ascaling_repulsive} show how the cross sections scale {\it relative to Born approximation} as a function of $A$.  Born approximation implies the following scaling of the nuclear cross section
\begin{equation}
\label{eq:Ascaling}
\sigma_{A}^{\rm Born}=\sigma_{p}^{\rm Born} \left(\frac{\mu_{A}}{\mu_{p}}\right)^{2} A^{2}~,
\end{equation}
where $\mu_{A}$ is the reduced mass of the $SA$ system.  This scaling fails badly even for $\alpha = 0.1$, for all cases: for both repulsive and attractive interactions and for both extended and point-like sources.   In the repulsive case, the true cross section is significantly lower than predicted by the Born approximation and the discrepancy increases with $A$.  There is no such simple relation in the attractive case, although at small \asn\ the true cross section is larger for the realistic extended case, while for large \asn\ the situation is more complicated. 

The peaks and valleys in cross section for the attractive case can be understood as follows, focussing on the point-like (exact Yukawa) case where the discussion is simple.   The Schroedinger equation for the Yukawa potential can be put into dimensionless form such that $\sigma \, m^2$ is a function only of the dimensionless parameters (in natural units, with $c=1$) \cite{buckleyFox10}:
\be
a \equiv \frac{v}{2\alpha} \,,
\quad
b \equiv \frac{2 \mu \alpha}{m} \, ,
\quad
x \equiv 2 \mu \alpha r\, ,
\quad
\tilde{V}(x) = -\frac{1}{x} e^{-\frac{x}{b}}\, ,
\ee
where as before $\mu$ is the DM-nucleus reduced mass and $m$ is the mediator mass.  Evidently, $b$ sets the range of the potential.
At low energy S-wave scattering ($l=0$) is dominant so the cross section is
\begin{equation}
\label{eq:cxtotSwave}
\sigma_\text{S-wave}=\frac{4\pi}{a^2 b^2 m^2}\sin^2(\delta_{0}) ~.
\end{equation}
When the S-wave phase shift $\delta_{0} \rightarrow \frac{\pi}{2}$, the cross section is on resonance and reaches its maximum value. The position of the peaks are in one-to-one correspondence with the zero energy bound states of the Yukawa potential.  When 
$b$ is small the potential is too narrow/weak to accommodate any bound states.  As $b$ increases, the potential becomes wider/stronger, up to the point where a bound state with $E_0 \rightarrow 0^-$ appears, at which value particle scattering has maximum cross section.  As $b$ continues to increase, the ground state binding energy gets more and more negative, up to some point where another bound state with $E_1 \rightarrow 0^-$ emerges and there is another peak in the scattering cross section. The position of these zero energy bound states are easily calculated to be at $b=1.68, 6.45, 14.34...$, which is exactly where the peaks in the cross section are located.  The physical implications of the bounds states that exist when the potential  is attractive is discussed in Sec. \ref{sec:bbnhybrids} below.

On the other hand when $\delta_{0} \rightarrow n\pi$ the S-wave cross section vanishes, which is an anti-resonance and corresponds to the dips in the cross section.  The small cross sections at the anti-resonances lead to gaps in the exclusion limits, as we shall see below. The anti-resonances are not associated with any bound state behavior;  they occur at $b=4.52,11.84, ...$.  

\begin{figure}
\centering 
\includegraphics[width=\linewidth]{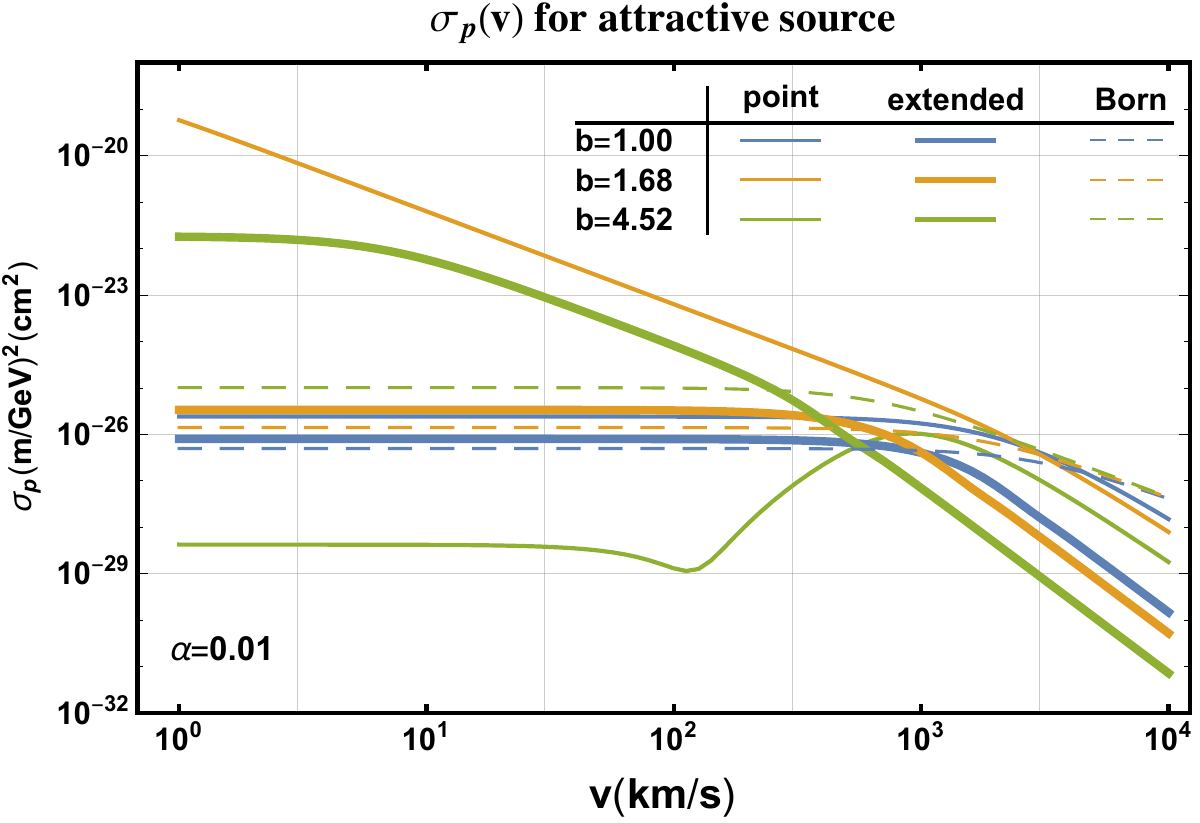}
\caption{\label{fig:Vdep} Velocity dependence of $\sigma_1$ for an attractive interaction with $\alpha=0.01$, for 3 values of $b$.  $b=1.68$ (ochre) is on resonance and $b=4.52$ (green) is on anti-resonance for the point-like case while the resonance and anti-resonance occur for different $b$'s for an extended source.}
\end{figure}

\begin{figure}
\centering 
\includegraphics[width=\linewidth]{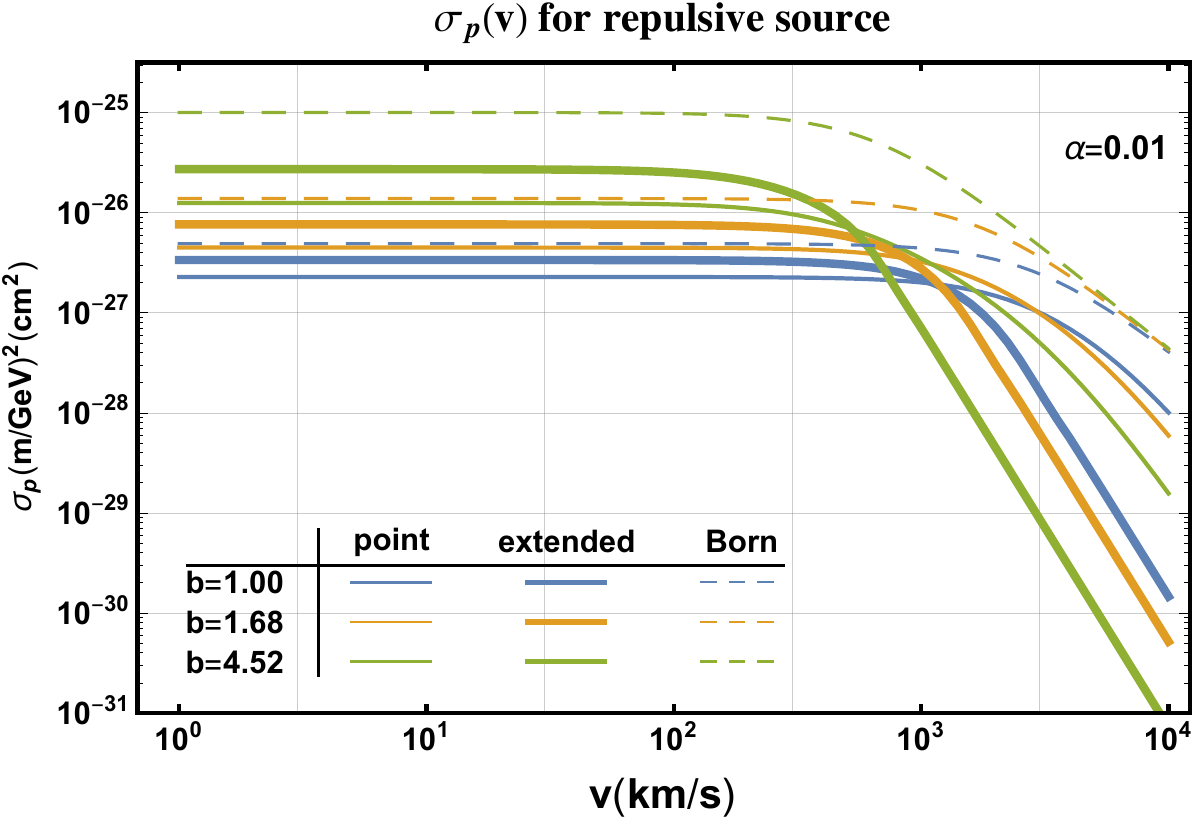}
\caption{\label{fig:Vdep_repulsive} As in Fig.~\ref{fig:Vdep}, for repulsive interaction.  Note the different vertical scale.}
\end{figure}

The existence of resonance and anti-resonance scattering is associated with non-trivial velocity dependence of the cross-section, as shown in Fig.~\ref{fig:Vdep}.  On resonance, $\sigma \sim v^{-2}$ at small velocity for a pure Yukawa, leading to the enhanced cross-section.  For generic values of $b$ the cross section is roughly constant for small $b$, but as the anti-resonance is approached ($b=4.52$ for point-like source), the cross section is diminished at small velocities.  At large velocities, for all $b$, $\sigma \sim v^{-4}$ and the scattering is Coulomb-like.   The transition to this regime occurs for $a> a_{\rm crit}(b)$.  When the source of the Yukawa is smeared over a nucleus, the behavior is different but still non-trivial.  
Fig. ~\ref{fig:Vdep_repulsive} shows the behavior for repulsive interactions, where there is no resonance or anti-resonance.

\section{Direct Detection and cosmological limits}
\label{sec:DDconstraints}

In view of the highly non-perturbative behavior seen in Figs.~\ref{fig:sig_alpha} - \ref{fig:Ascaling_repulsive}, we must re-examine the existing limits on DM-nucleon interactions.  The limits on spin-independent cross sections are almost universally reported as bounds on the DM-nucleon cross section.   However in almost all cases, what has actually been done is to use Born approximation and the $A$-scaling it implies to convert experimental limits on $\sigma_{A}$ to the reported limit on $\sigma_{1}$.  The cross section is usually assumed to be independent of $v$, or in some analyses various power-law dependences are explored.

Here, we report excluded regions in the $\alpha_{SN}$-$m_S$ (more generally $\alpha$-$m_{DM}$) plane, which is the most fundamental information, rather than reporting limits on $\sigma_{DMp}$.  For a given $\alpha$ and DM particle velocity and mass,  we use the correct value of the cross section from high-resolution tabulated numerical solutions to the Schroedinger equation.  This naturally incorporates the actual velocity dependence which as seen above is not in general a simple power-law at all velocities.  We generally report limits taking $m$ = 1 GeV.   

\subsection{Correcting the XQC limits} 
\label{sec:XQC}
The X-ray Quantum Calorimeter~\cite{xqc} was an experiment to measure the diffuse X-ray background using microcalorimeters onboard a sounding rocket at about 100 km altitude in the atmosphere. The results can also be used to put limits on the DM-nucleon cross section, and extensive studies have be performed~\cite{zfWindowDM05,erickcek+07,mfWindow17,mfVelDep18}.  The analyses prior to \cite{mfVelDep18} all assumed that the entire recoil energy of the Si nucleus ($\sim$ keV for DM mass of 2 GeV) is converted to phonons in the XQC calorimeter.  However as pointed out in \cite{mfVelDep18}, at these low energies below ionization threshold, the recoiling Si atom moves as a whole and produces a cascade of dislocated atoms -- very low energy (few eV) lattice defects called Frenkel pairs consisting of a hole in the lattice and an interstitial Si atom~\cite{BarYamInterstitialPRL84,Tang+Interstitial97,Leung+InterstitialPRL99,Huhtinen02,srour+DisplacementDamageReview03,Rinke+InterstitialPRL09,Gusakov+FrenkelPair09,Junkes11}.  Similar effects can be expected in sapphire and other materials used for micro-calorimeter detectors~\cite{sapphire09}.  So rather than being thermalized, the energy deposited by the DM collision may be stored in Frenkel pairs.  Limiting the production of such point defects (vacancies and interstitial atoms) during the process of crystal growth, is an important commercial in the semi-conductor industry.  For a review see \cite{vanLintRadiationBook80}.  A typical concentration of defects is $10^{15} \,{\rm cm}^{-3}$;  compared to this, the increment in defects during the course of the 100 s XQC mission is negligible, as we now show.  

The Galactic DM abundance in the solar neighborhood is $\approx 0.3\, {\rm GeV \,cm}^{-3}$.  Taking $m_S = 2 m_p$ and DM velocity $\approx 300$ km/s, the flux of \sdm\ is $\approx 10^{6} \,{\rm cm^{-2} s^{-1}}$.  The fraction of the DM kinetic energy deposited per scattering is 
\be
f_{\rm KE} = 2 (1 - \cos \theta) \frac{m_{\rm DM} m_{\rm A} }{(m_{\rm DM} + m_{\rm A})^2} ,
\ee
where $\theta$ is the CM scattering angle.  For $m_{\rm A} >> m_{\rm DM}$, appropriate for \sdm\ and detectors like XQC, and noting that the scattering is isotropic in the CM, the average kinetic energy deposited in the initial collision is 
$\approx 100$ eV, thereby producing $\lesssim 100$ point defects.   Even if every DM particle scattered once (vastly more than consistent with expectations and other constraints), the increment in the density of defects in a 3 mm thick Si detector in the 100 s flight would be $\lesssim 10^{11} \, {\rm cm}^{-3}$ -- more than 4 orders of magnitude less than the initial density of defects and hence not noticeable at all.   Likewise the impact of DM on the abundance of point defects at sea-level is insignificant: if the scattering cross section is large, the ambient DM at sea level carries little energy and a scattering produces few or no defects, whereas if the cross section is small, DM-scattering is a minor contributor to the point-defect population.  Perhaps some detection strategy could make use of this process, however.  

Due to the likelihood that recoiling atoms deposit their energy in creation of Frenkel pairs rather than in thermal excitations of the material, until the thermalization efficiency of detectors such as XQC, DAMIC and CRESST is measured with neutron scattering, limits from those experiments can only be discussed in terms of their potential sensitivity, not as actual exclusion limits.  

\begin{figure}
\includegraphics[width=\linewidth]{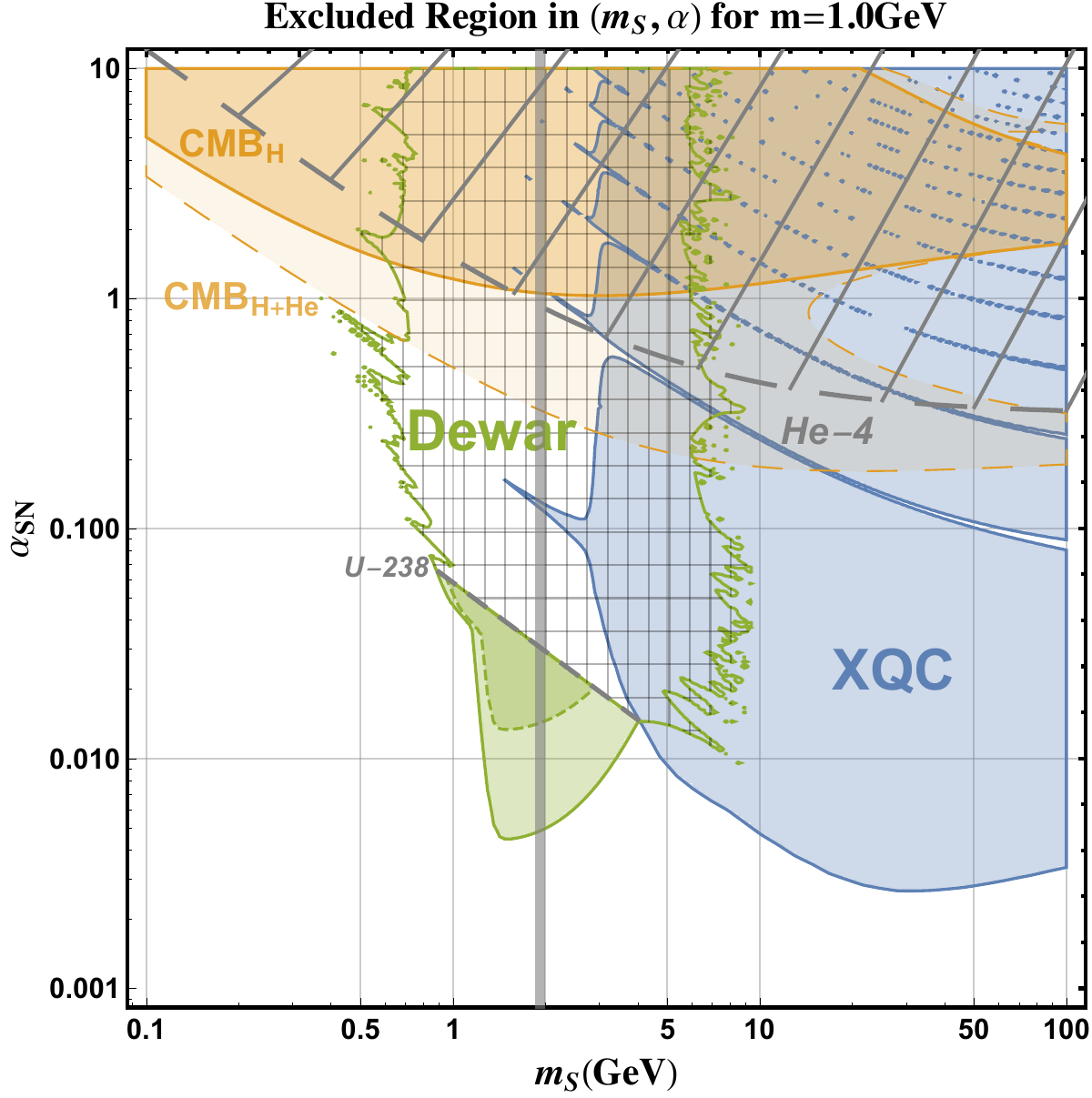}
 \caption{Excluded regions in the $\alpha_{SN}$-$m_S$ plane from XQC (blue), CMB (tan) \cite{xuDvorkin+18} and dewar experiments (green\&checkerboard), for an attractive $SN$ interaction and uniform sphere nuclear distribution with $R_0 = 1.0\,$fm.  (If the DM-nucleon interaction is repulsive, the plot is similar except that the XQC and dewar boundaries are smoother and the dewar limits exclude the entire green-bounded region; see \cite{xf20}.)  The thin vertical grey band marks the range $m_S = 1860-1890$ MeV.  The diagonal-hatched region above the upper grey-dashed line is excluded by the requirement that primordial $^4$He not hybridize with DM.   Within the checkerboard region the dewar limits are inapplicable when the interaction is attractive, because DM would hybridize with nuclei in Earth's crust.  The darker green region with dashed boundary shows the dewar exclusion in case the local DM density is a factor-10 lower than calculated by \cite{nfm18}, to give an indication of the sensitivity.  The light tan region bounded by dashed lines is a potential exclusion region estimated from the CMB analysis of \cite{xuDvorkin+18}, approximately taking into account ``resonant" DM-\He\ scattering.    
  }\label{fig:DDexclusions}
\end{figure}

The blue shaded region of Fig.~\ref{fig:DDexclusions} shows our exclusion region in the $\{\alpha, m_{\rm DM}\}$ plane from XQC data, for a mediator mass of 1 GeV.  We adopt the fiducial thermalization efficiency $\epsilon_{\rm th} = 0.01$; this is at best an indicative, round-number value motivated by a Frenkel-pair cascade analysis  \cite{mfVelDep18}.  We follow the XQC analysis of \cite{mfVelDep18} but solve the Schroedinger equation for the extended Si nucleus to obtain the cross section, rather than using the $A$ scaling from Born approximation and the Helm form factor.  
(It should be noted that the Helm-form factor gives a poor approximation to the impact of an extended source, so replacing the Born approximation is not enough; see~\cite{xf20} for details.)   Future analyses of direct detection experiments aiming to detect DM under an overburden will also need to take into account the possibility of DM capturing on a nucleus en route to the detector, as discussed in Sec.~\ref{sec:hyb} below.  

\subsection{Limits from dewar experiments} 
\label{sec:dewar}
It was pointed out by Neufeld et al.~\cite{nfm18} that if the DM cross section on nuclei in the Earth's crust and/or atmosphere exceeds $\approx 10^{-28.5} \, {\rm cm}^{2}$, a DM atmosphere forms around the Earth; the density distribution of this atmosphere was characterized under the assumption that the atmosphere of the Earth has been reasonably steady over its lifetime.  The density of DM particles near the surface of the Earth reaches $\approx 10^{14} \, {\rm cm}^{-3}$ for $m_{\rm DM} \approx 2 m_p$.  The presence of such a high density of DM, if it is free rather than bound to nuclei, produces detectable effects that Ref.~\cite{nfm18} used to place limits on DM interactions with various materials.  The limits included heating of liquid He and other cryogens in dewars, HST orbital drag, and modification to the thermal conductivity of the Earth.  Stronger upper limits on $\sigma_{A}$ were obtained for many more individual $A$ values by dedicated dewar experiments in which samples of different $A$ were placed in a dewar of liquid nitrogen and the evaporative mass loss after a time interval was measured~\cite{nbn19}.   

Interpreting the limits on $\sigma_{A}$ for the ensemble of $A$'s is challenging due to the non-trivial dependence of $\sigma_A$ on $\alpha$ as discussed in the previous section (Sec. \ref{sec:DMints}).  Basically it entails stepping through the full parameter space, determining  at each point in parameter space the local number density based on the DM scattering cross section with air or crust including capture and evaporation from Ref.~\cite{nfm18}, then calculating the corresponding heating due to DM collisions for pellets of the given $A$.  (We thank D. Neufeld for providing necessary input data.)  Details of how this was done are presented in~\cite{xf20}.  

The green-bounded region in Fig.~\ref{fig:DDexclusions} is the nominal resultant exclusion region for an attractive DM-nucleon interaction.  The boundary in the repulsive case is similar but smoother; a plot for the repulsive case can be found in~\cite{xf20}.  
To get an estimate of the uncertainty in the excluded region from the dewar experiments and obtain a more conservative limit from the dewar experiments, we can take a factor-10 lower \sdm\ atmosphere than estimated in Ref.~\cite{nfm18}.  Such a reduction might arise from a lower Galactic DM flux or possible disruptions in the accumulation of a DM atmosphere.  This reduces the size of the excluded region below the U-238 line as indicated by the darker green region;  for $m_S = 2 m_p$ the limit on \asn\ becomes 0.015. 

Note, however, that when the interaction is attractive and \asn\ $\gtrsim 0.04$, the dewar exclusion shown Fig.~\ref{fig:DDexclusions} is evaded.  In that case, \sdm\ in the Earth binds to nuclei as discussed in Sec.~\ref{sec:hyb}.  DM particles bound to nuclei cannot freely penetrate through the walls of the dewar and heat the material, so the dewar limits on DM interactions do not apply; all other limits of \cite{nfm18} are evaded as well.   The cross-hatched region shows the part of the XQC exclusion region that is eliminated for this reason.  


\subsection{Limits on DM-baryon interactions from structure formation} 
\label{sec:CMB}
The cross-hatched region above the upper grey line is excluded for an attractive interaction because primordial He would hybridize with \s's shortly after being produced.   The region labeled CMB follows from the limits on $\sigma_{\rm DM-p}$ obtained in Ref.~\cite{xuDvorkin+18} from the damping of structure formation that results from DM-p interactions, translated to $\alpha,\,m_{\rm DM}$ space taking the proton to be a hard sphere of radius 1.0 fm.  The lighter region bounded by dashed lines and labeled CMB$_{H+He}$ is derived from limits of \cite{xuDvorkin+18}, but instead of assuming a fixed ratio $ \sigma_{S \rm He} /\sigma_{S \rm p} $ based on Born approximation (their Eq. (14)) and a coupling proportional to charge, we used the actual ratio determined by each $\alpha,\,m_{\rm DM}$ pair and the Schroedinger equation.  
For \asn\ such that He scattering is in the ``resonance" region, He completely dominates H in its drag on DM.  But recombination of He occurs at higher $z$ than for protons, so the rough estimation employed here probably exaggerates the limit.  We present it here because this is an interesting region of parameter space in connection with exotic isotopes and deserves a dedicated cosmological study along the lines of~\cite{xuDvorkin+18}. 

The recent limits from \cite{ngbw19} are not directly applicable to \sdm\ because non-gravitational DM-baryon interactions modify the structure of cores of dwarf galaxies, potentially modifying the mapping from as-observed dwarf properties to linear-regime structure used in \cite{ngbw19}; another issue which has not yet been examined by the community is the extent to which uncertainties due to the complexities of baryonic physics have a significant impact on the analysis of \cite{ngbw19}. 

\subsection{Astrophysical Limits}

Ref.~\cite{wf19} reports limits on dark matter-ordinary matter cross sections for a number of dark matter scenarios, from bounds on the heating or cooling of the Leo T dwarf galaxy and Milky Way gas clouds due to DM scattering on nucleons and electrons.  The strongest limits for DM-nucleon scattering for $m_{\rm DM} \approx 2 m_p$ come from the gas clouds, whose limit is included in Fig.~\ref{fig:DDexclusions}.  It is quite similar to the CMB limit for the $m_{\rm DM} \approx 2 m_p$ region of interest for \sdm.  Given the entirely different uncertainties entering the cosmology and astrophysics analyses, it is useful to have both.  The limit from Ref.~\cite{wf19} is similar to the limit from requiring that primordial $^4$He does not form bound states with \sdm, but applicable for either attractive or repulsive interactions.
 
\begin{figure}
\centering 
\includegraphics[width=\linewidth]{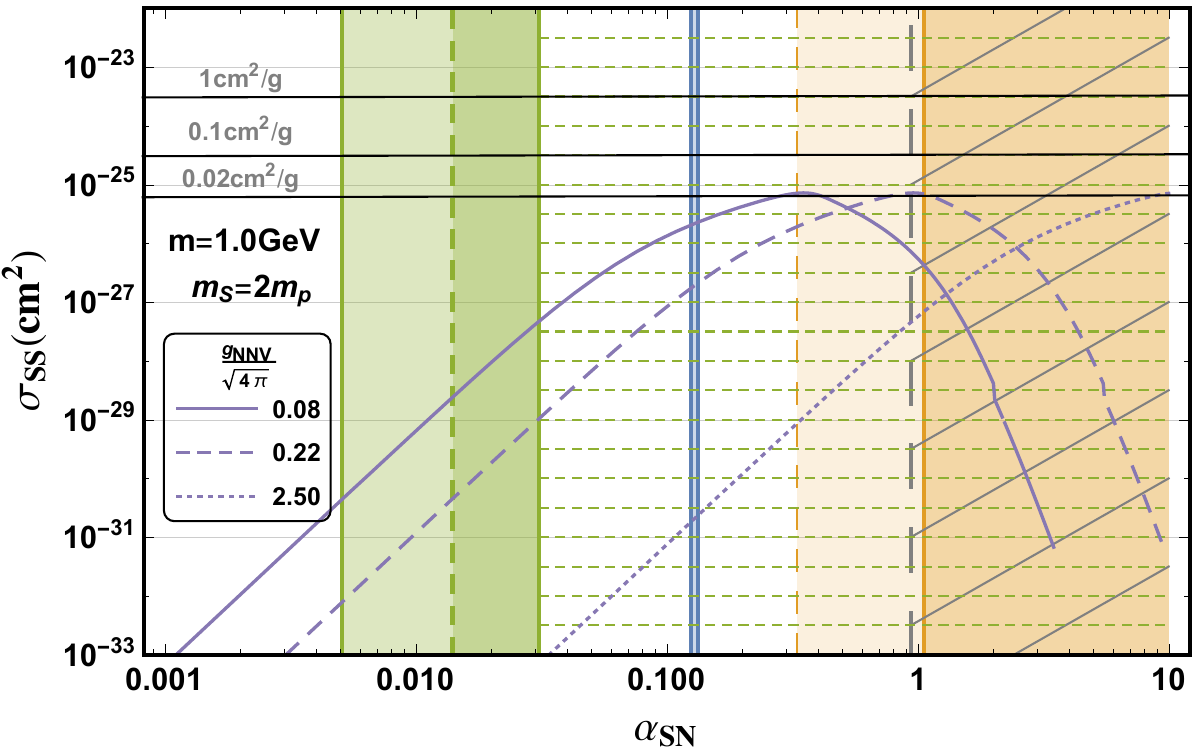}
\caption{\label{fig:sigSS} The SIDM cross section as a function of \asn\ taking $m_S = 2 m_p$ and $m= 1$ GeV, for $g_{NNV}/\sqrt{4 \pi} = 2.5, 0.22$ and 0.08, with the key limits from Fig.~\ref{fig:DDexclusions}.  The green range 0.015 $<$ \asn $\lesssim 0.03$ is excluded by dewar limits (less conservatively, 0.005-0.03) while the tan region is excluded by the CMB.  The pale tan is the estimated potential exclusion region by extending the CMB analysis to $^4$He with non-perturbative interactions as discussed in the text.  The horizontal black lines mark $\sigma_{SS}/m_{\rm DM} \sim 1,\,0.1$ and $0.02 \, {\rm cm^2/g}$, suggesting that sexaquark dark matter does not have sufficiently strong self-scattering cross section to have a significant astrophysical impact.}
\end{figure} 

\section{Self-Interacting Dark Matter}	
\label{sec:SIDM}
Spergel and Steinhardt \cite{ss:SIDM} pointed out two decades ago that DM self-interactions of order $\sigma/m_{\rm DM} \sim 1\, {\rm cm^2/g} \,( = 1.78 \times 10^{-24}\, {\rm cm^2/GeV}$) could explain the observed cores in galaxy centers, whereas cosmological simulations predicted cusps.  More realistic treatment of baryonic physics since 2000 alleviates the most severe problems of LCDM, so the call for self-interacting DM (SIDM) is less pressing.  However problems with LCDM continue to be identified, e.g., \cite{TooBigToFail11}, and several authors argue that a much better accounting is obtained of galactic dynamics with self-interactions of order $\sigma/M \approx 1 \,{\rm cm^2/g}$.  The literature is very large; see \cite{weinberg+SIDM15,creasey+17,SIDMrotCurvesPRX18} for examples of recent papers and \cite{TulinYuPhysRpts17} for  a review.  For the characteristic \sdm\ mass $m_{\rm DM} = 2 m_p$, $\sigma/M \approx 1 \,{\rm cm^2/g}$ translates to $\sigma_{SS} = 3.34 \times 10^{-24}\, {\rm cm^2}$, at velocities $\approx 100$ km/s relevant to the small scale structure of galaxies.  

\sdm\ has self-interactions via exchange of whatever produces its interactions with nucleons, identified here as predominantly due to the exchange of the flavor-singlet vector meson denoted here by $V$ (Sec. \ref{sec:DMints}).  Taking the \s\ interaction to be dominated by $V$ exchange, the self-interaction is repulsive because like charges repel for vector exchange.  The Yukawa interaction has the same form as in Eq.~\ref{eq:yukawa}, with
\be
\label{aSS}
\alpha_{SS} = \left(\frac{\alpha_{SN}}{g_{NNV}/\sqrt{4 \pi} } \right)^2 = 0.16 \,  \left( \frac{2.5 }{g_{NNV}/\sqrt{4 \pi} } \right)^2 \alpha_{SN}^2~,
\ee
where $g_{NNV}/\sqrt{4 \pi} = 2.5$ is taken from the Nijmegen Soft Core NSC89 its (Table VI of \cite{NSC89}).   Note, however, that although low energy hadron interactions have been modeled in detail, there is a factor-few uncertainty in the estimate (\ref{aSS}) because we ignore here the derivative ($f$) coupling of the $V$ and because the $V$ couplings obtained with the more elaborate ESC19 Extended Soft Core model \cite{ESC19} differ from those from NSC89, especially the $f$ couplings.  Although we drop those contributions, they can interfere and cause significant cancelations or enhancements.   

Figure \ref{fig:sigSS} shows the $SS$ cross section as a function of \asn, with $m_V = 1$ GeV and several choices for $g_{NNV}/\sqrt{4 \pi} = 2.5, \, 0.22$ and 0.08.  Exclusion regions from the previous section if the \s-nucleon interaction is attractive are also indicated.   The cross-hatched region is excluded by the requirement that sexaquarks do not bind to \He.  The orange region is excluded by our analysis based on the CMB limits on $\sigma_{\rm DM-p}$ of \cite{xuDvorkin+18}, and the pale-orange region by the estimated extension to \He\ discussed in Sec. \ref{sec:CMB}.   
The green region is excluded by the ~\cite{nfm18,nbn19} dewar limits and the blue region by XQC.  The horizontal lines mark $\sigma/m_{\rm DM} \sim 1,\,0.1,\, 0.02\, {\rm cm^2/g}$ taking $m_S = 2 \, m_p$.  One sees from Fig.~\ref{fig:sigSS} that the maximum value $\sigma/m_{\rm DM} \approx 0.02\, {\rm cm^2/g}$, which is lower than usually considered by advocates of SIDM to improve the description of galaxies relative to LCDM.  The maximum value may be lower still, depending on \asn\ and $g_{NNV}/\sqrt{4 \pi} $.  Given that \sdm\ also interacts with baryons, it is possible that considering both self- and baryonic interactions, \sdm\ can give as successful a description of galaxies as claimed for SIDM, with a lower self-interaction cross-section than needed in pure SIDM.


\section{Production of exotic isotopes}
\label{sec:bbnhybrids}

\begin{figure}
  \includegraphics[width=\linewidth]{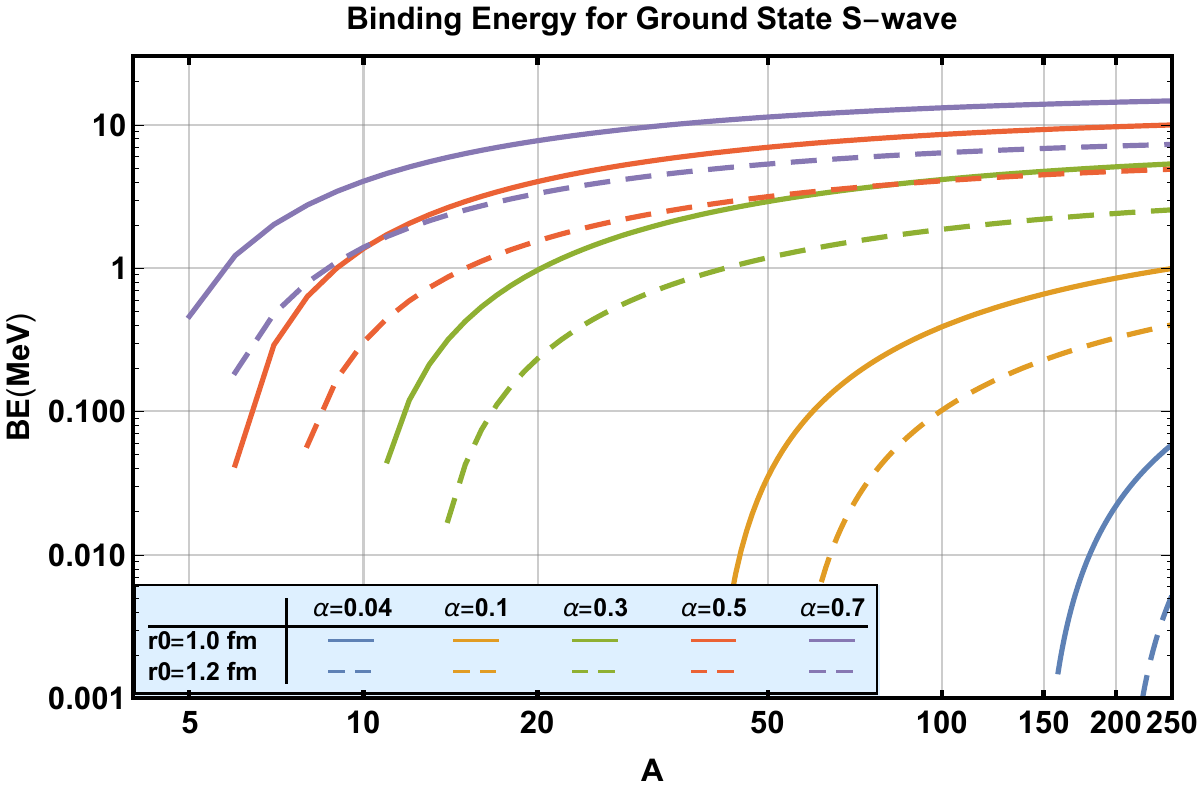}
 \caption{Binding energy of a sexaquark as a function of atomic mass for selected \asn; solid (dashed) curves correspond to nuclear radius parameter $R_0 = 1.0 \,(1.2)\,$fm.  Note that the allowed range of \asn\ shown in Fig.~\ref{fig:DDexclusions} for $R_0=1.0\,$fm shifts for $R_0=1.2\,$fm as discussed in more detail in \cite{xf20}. }  \label{fig:BE}
\end{figure}

DM-nucleus hybridization takes place at a phenomenologically interesting level,  if a bound state exists and the relative velocity is sufficiently low for capture to occur, in regions of the Universe with sufficiently high density of both \sdm\ and nuclei of mass $A \geq A_{\rm min}(\alpha_{SN})$.   Fig. \ref{fig:BE} shows that the binding energy of a sexaquark to a nucleus of mass $A$ is typically $\approx$ 0.1-10 MeV over the range of \asn\ large enough to produce binding for $A \lesssim 40$.  Therefore most bound states would survive  
in most astrophysical environments, once produced.   In this section we consider three particularly relevant production sites, Big Bang nucleosynthesis, stars and in the Earth.

Avoidance of creation of abundant (and not observed) primordial $^4$He$S$ implies \asn\ $\lesssim 0.7$ as discussed in Sec.~\ref{sec:BBN} below, which has the further consequence that sexaquark binding only occurs at a significant level with nuclei produced in stars and their explosions and mergers.   This means that non-negligible levels of hybrid nuclei is restricted to baryon-dominated environments, so that depletion of free \sdm\ by binding to baryons has only a higher-order impact on the CDM character of \sdm. 

Denoting the hybrid nucleus-sexaquark state by $AS$, the dominant capture process is $S + A \rightarrow AS + \gamma$, analogous to the case for neutron capture.   The theory of neutron capture is well-developed due to its importance in nucleosynthesis and nuclear physics, c.f.,  \cite{Lynn}.  At intermediate to high incident energy, resonance intermediate states can be excited, but at the low energies relevant here, the most important contribution to the amplitude is due to ``direct capture" with the 
nucleus a passive source of the potential.  
Because the potential well is shallower in the $AS$ case than for nuclei, given the \asn\ range of interest, the dominant contribution to the capture cross section is from an initial p-wave scattering state into the (s-wave) bound state with emission of an E1 photon.  Derivation of the capture cross section in this case will be presented elsewhere~\cite{fxe20}.   

\subsection{Primordial nucleosynthesis}
\label{sec:BBN}
\sdm\ can potentially impact primordial nucleosynthesis in two ways: \\
\noindent{ 1)} interference with the standard synthesis process by disruption of intermediate or final nuclei, and/or  \\  
\noindent { 2)} production of hybridized nuclei, if the interaction is attractive and sufficiently strong to form a bound state.

Our colleague R. Galvez modified the BBN code AlterBBN \cite{alterBBN17} to allow for \sdm-nucleus interactions.  He found that even with the DM-nucleus cross section at its unitarity limit, nuclear breakup by \sdm\ scattering ($A+S \rightarrow A_1 + A_2 + S$, e.g., $^7Be + S \rightarrow ^3He + ^4He + S$) is negligible.  This stems from the kinematic fact that in the temperature range at which nucleosynthesis occurs, $T \lesssim 80$ keV, {\it i) } the available energy in the final state is so small that 3-body phase space suppression is very large and {\it ii)} the energy transfer in collisions between nuclei and \sdm\ is too small to overcome the potential barrier to breakup except in the extreme tail of the Boltzmann distribution, even for the least-bound nucleus \Be.   

However as noted in Sec. \ref{sec:DMints},  if the \s-nucleon interaction is attractive,  \s\ forms bound states with nuclei of $A>A_{\rm min}(\alpha_{SN})$.  The minimum nuclear mass to form an $A$-\s\ bound state increases as the coupling strength \asn\ decreases.  The relationship is shown in Fig. \ref{fig:Amin}.  
It is very important to stress that all $A$ above some minimum value $A_{\rm min}$ which depends on \asn\ will form bound states.   If nucleus $A$ hybridizes, so do all heavier nuclei, in the approximation of spherical homogeneous nuclei with radii increasing as $\sim A^{1/3}$.

For the $\alpha_{SN}$ range of interest, hybridization occurs at a rate much higher than the expansion rate of the Universe at the BBN epoch ($\tau_{\rm Univ} \approx 100$ s).  Thus hybridized and unhybridized states are in chemical equilibrium.  Unless the binding energy is very small, statistical physics favors hybridization so we require $\alpha_{SN}$ to be small enough that no essential primordial nuclei forms a bound state with \s.  This constrains $\alpha_{SN} \lesssim 0.7-0.8$ in the attractive case, to avoid that primordial He is not in fact He$S$ with mass $\approx 6$ amu.  As noted, this condition also insures that no other nucleus having $A\leq 4$ can hybridize.  

An enticing scenario for explaining the deficit of primordial \li\ as inferred from the Spite plateau \cite{fieldsBBN11,pdgBBN17,asplund06,sbordone+10,cocVBBN17}, would be if \Be\ hybridized such that the electron capture production of \li\ or $^7{\rm Li}S$ was suppressed or forbidden.  While this can be contrived \cite{xf20}, it does not solve the problem because then $^7{\rm Be}S$ should be present at the level expected for \li, whereas the upper limit on Be in dwarf stars is much lower~\cite{molaroBe1984,spiteBe19}.   (The isotope shift for 2 amu higher mass is small relative to the widths of the lines used to identify Be or Li, so a hybridized component in the elemental abundance would not be missed.  We have not explored whether hybridization of nuclei with $A>4$ during BBN could diminish production of \li\ and resolve the \li\ puzzle.)  

\subsection{Exotic isotope formation in Earth and its atmosphere} 
 \label{sec:hyb}

Dedicated studies place very stringent limits on the local abundance of exotic isotopes with $\gtrsim 100$ GeV mass splittings relative to the normal nucleus~\cite{mullerAlvarez+77,smithBennettAnomH79,smith+AnomH81,hemmick+ExoticIso90,javorsek+Fe&AuPRL01,javorsek+Gold01,mueller+AnomIsoHe04}.  However limits have yet to be developed for mass splitting close to 2 amu, as is relevant for the exotic isotopes formed if \sdm\ hybridizes with nuclei.  In this section we take a first look at possible exotic isotope abundances for attractive \sdm.   Many of the most important nuclei in the Earth have stable isotopes with $A+2$, with natural abundances generally much larger than expected for \sdm\ hybrids, so dedicated experiments to search for the predicted isotopes will be required for most cases, to discriminate between an $A+2$ nucleus and an $AS$ bound state.   

\begin{figure}
\centering 
\includegraphics[width=\linewidth]{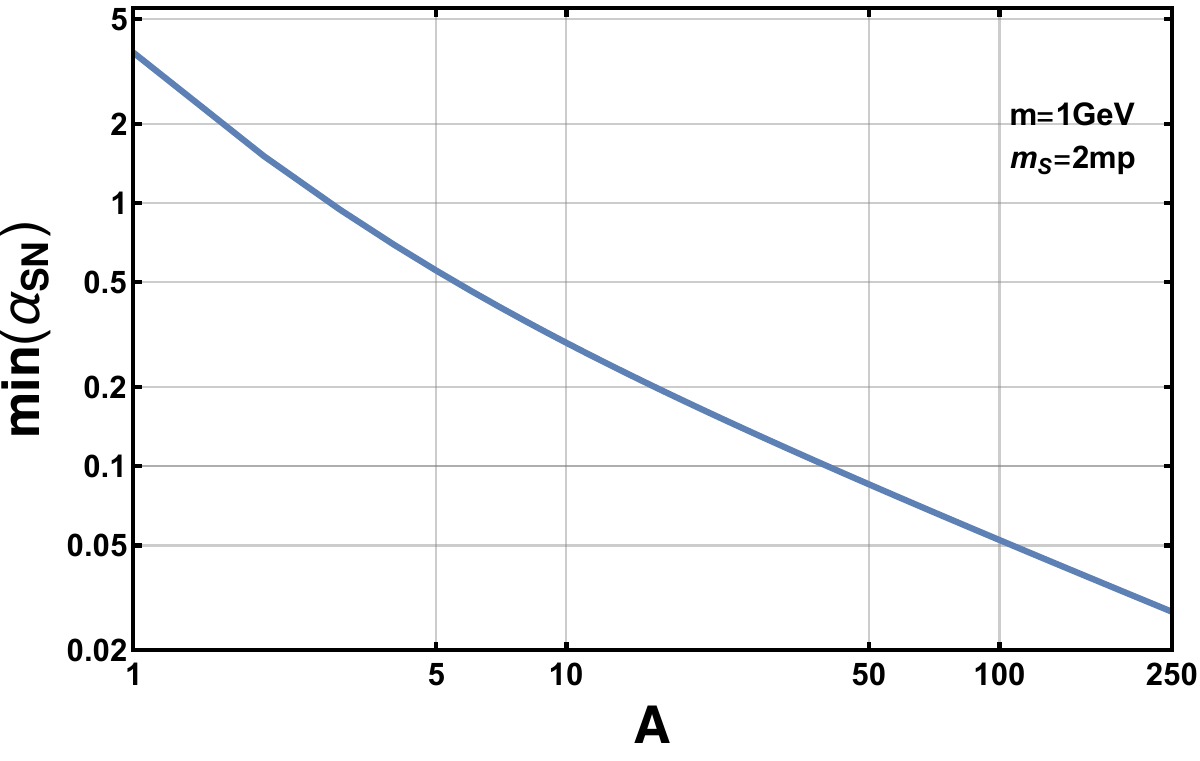}
\caption{\label{fig:Amin} The minimum \asn\ such that a bound state $AS$ can form; \asn\ $\gtrsim 0.7$ is excluded by non-binding of primordial $^4$He.  Mentally inverting axes, the plot also gives $A_{\rm min} (\alpha_{SN})$, the minimum $A$ which forms bound states for given \asn.}
\end{figure} 

As shown in \cite{f:144}, about 15\% of hadronically interacting DM (HIDM) with mass $\mathcal{O}$(GeV), whose trajectories intersect Earth, are captured.  Ref. \cite{nfm18} develops in detail the theory of the HIDM atmosphere which forms around Earth, in the absence of hybridization.   The number of \sdm\ particles captured by Earth over its lifetime can be estimated in different ways.  Here we use $N_S \approx 1.3 \times 10^{41}$ as calculated in ~\cite{nfm18}, for an accumulation rate of 
\be
\label{accum}
\dot{N}_{S} \approx 2.8 \times 10^{37} ~ {\rm Myr}^{-1}~.
\ee 
$\dot{N}_{S}$ can arguably be up to a factor-6 higher~\cite{fxe20}, increasing proportionately all of the abundance estimates below. Here we give a very simplified treatment of the hybridization process, for some illustrative possibilities.

In a collision with the relatively massive nuclei comprising the Earth and its atmosphere, an energetic DM particle of mass $2 m_p$ loses 4/9 of its energy, on average.  To go from $v_{\rm esc} =$ 11 km/s after the DM is gravitationally bound, to $v\approx$ 1 km/s for thermal equilibrium at 300 K, requires of order 10 collisions -- negligible compared to the total number of scatterings typically preceding a hybridization event.  
The mass of the DM is small relative to the nuclei it scatters on, and the velocities are very small, so the CM and lab frames are essentially the same and the scattering is isotropic in the lab frame.  

As SDM diffuses through the atmosphere or in the Earth, two processes are at play:  

\begin{enumerate}

\item The SDM {\it scattering length} is determined by its interaction with {\it all} of the nuclei in the local environment : 
\be
\label{scatlen}
{\bar{\lambda}} = \left(n\, \sum_A f_A \, \sigma_A~\right)^{-1}~,
\ee
where $f_A$ is the fractional abundance by number of nucleus $A$ and $n$ is the total number density.  The total path length $L$ while traversing a thickness $z_0$ is then, in homogenous diffusion approximation:
\be
\label{pathlen}
L  = 3 \, z_0^2/ \bar{\lambda} = \bar{\lambda} \, N_{\rm scat}  ~,
\ee
where the last relationship enables the mean number of scatterings in the medium to be deduced.  
\item  The SDM {\it capture length} is determined by its capture cross section on those nuclei with $A \geq A_{\rm min}$ capable of forming bound states ($A_{\rm min}$ depends on \asn, as shown in Fig.~\ref{fig:Amin}), giving:
\be
\label{caplen}
\bar{\lambda}^{\rm cap} =  \left( n \, \sum_{A \geq A_{\rm min}} f_A \,\sigma^{\rm cap}_A   \right)^{-1}~.
\ee

\end{enumerate}
Capture occurs if $L\gtrsim \bar{\lambda}^{\rm cap}$.  

\begin{figure}
  \includegraphics[width=\linewidth]{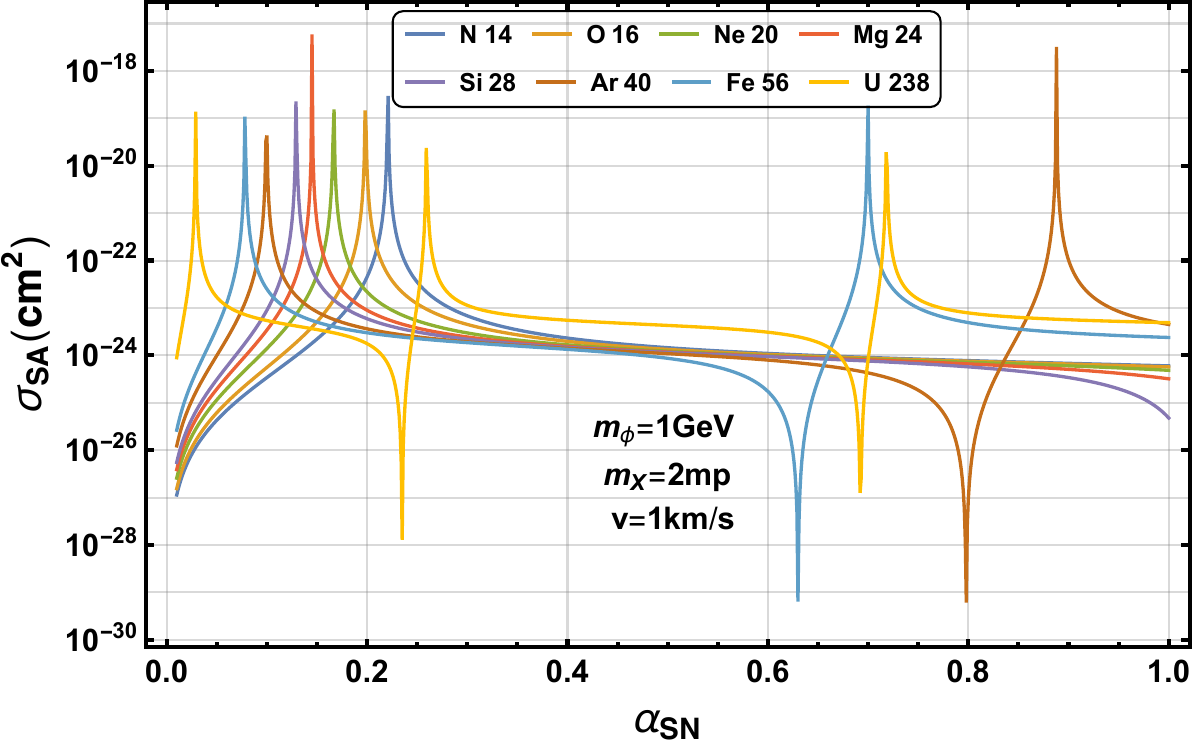}
 \caption{Cross section in cm$^2$ versus \asn\ for selected nuclei  \{$^{14}$N, $^{16}$O, $^{20}$Ne, $^{24}$Mg, $^{28}$Si, $^{40}$Ar, $^{56}$Fe, $^{238}$U\}, for an attractive interaction.  Below \asn\ = 0.03 there are no bound states for $A \leq 238$.   The positions of the peaks scale approximately linearly in the nuclear radius parameter $R_0$  
 }  \label{fig:sig8}
\end{figure}

Fig.~\ref{fig:sig8} shows the scattering cross section versus \asn\ for a number of interesting or abundant elements in the Earth's crust.  For \asn\ $\geq 0.03$ (the minimum to bind $^{238}$U), all \sdm\ captured by Earth become bound to nuclei and the dewar limits discussed in Sec.~\ref{sec:dewar} (and indeed all limits of \cite{nfm18}) are evaded.  However the distribution over host elements and physically in the Earth, of the $\approx 1.3 \times 10^{41}$ \s's accumulated over the Earth's lifetime, varies dramatically with the value of  \asn.   

The simplest scenario to analyze is when DM predominantly captures on O in the ocean, which we discuss to illustrate the analysis; a more comprehensive study of other possibilities is presented in ~\cite{fxe20}.  Natural O is predominantly $^{16}O$, with 0.2\% $^{18}O$.  Since the scattering and capture cross sections depend on $A$, not $Z$, these are not exactly equivalent regarding their capture probability, but we lump them together in this first, broad-brush description.   Moreover the O in the atmosphere (which is mostly in $O_2$) rapidly cycles with O in the oceans and biomass on the timescales relevant to this problem, and furthermore the column number density of O in the atmosphere is small compared to that in $H_2O$ in the ocean, so we lump together O from both atmosphere and ocean, for this simple treatment. 

The oceans cover 70\% of the Earth's surface, with mean depth 5 km and scattering length $\lambda =  30 \,{\rm cm} / \sigma_b \,$, for $N_{\rm scat} \approx 10^{12}$ from Eq. (\ref{pathlen}).  Thus \sdm\ incident on the surface of the oceans will be captured by O in the ocean water, as long as an OS bound state exists and $\sigma^{\rm cap}_O/ \sigma_O \gtrsim 10^{-12}$.  (For comparison, neutron capture cross sections are typically $10^{-4} - 10^{-5}$ times the scattering cross section; if this ratio were applicable to \sdm, capture would take place after $\approx 10^{4-5}$ collisions, or capture depth $z^{\rm cap} =\sqrt{\lambda \lambda^{\rm cap}} \approx 30\,$m.)  Ocean water has a residence time of $\approx 100\,$Myr, during which $0.7 \times 2.8 \times 10^{39} $ \sdm\ particles capture.  The total number of O nuclei in the oceans and atmosphere is $5.5 \times 10^{46}$ for a fractional abundance of OS relative to all O of  $\approx 3.6 \times 10^{-8}$.  

If the capture to scattering cross section ratio is smaller than $10^{-12}$, capture would not occur on first passage through the ocean and the $OS$ would be distributed more uniformly through the solid Earth, since O is the most abundant element in rocks.  Assuming $OS$ bound states exist and O is the dominant binding site for the \sdm, a lower limit on the fractional abundance is given by assuming it is distributed uniformly over all the O in the Earth:
\be
\frac{N_{OS}}{N_O} \gtrsim \frac{1.3 \times 10^{41}}{6.7 \times 10^{49} } \approx 2 \times 10^{-9}~.
\ee
Comprehensive predictions of the distribution of captured \sdm\ over different molecules and physical locations in Earth and its atmosphere, for general \asn, will require a multi-disciplinary effort.  It goes without saying that the estimates of this section are factor-two level at best.

\subsection{Production of $AS$ bound states in stars} 
 \label{sec:stars}
Potentially adding to the exotic isotopes created in Earth and Moon, are those created in the previous generations of stars and distributed via supernovae into the material from which the solar system formed.   In this subsection we give a very rough, preliminary estimate of whether stellar production could be a significant source of local exotic isotopes or not.   

Virtually all SDM particles that encounter a stellar surface become gravitationally bound.  Stars have a density similar to water, so the scattering  and capture length discussion of the previous subsection applies and all the SDM will be bound into exotic isotopes if there are nuclei present of $A\geq A_{\rm min}$.  In the approximation that the star is homogeneous, the relative abundances of exotic isotopes is in proportion to the capture cross section for element $A$.  In the approximation that the capture cross sections are equal, the fraction of exotic isotopes will be equal for all $A\geq A_{\rm min}$, and equal to the total number of accreted \s's divided by the total number of nuclei with $A\geq A_{\rm min}$.  

Following the calculation leading to Eq.~\ref{accum}, dividing by 0.15 since all SDM particles are captured and rescaling by $(R_\star/R_{\rm Earth})^2$, gives $N_S \approx 10^{46}  $ for a star of solar radius  accumulating for 5 Gyr.   Taking the star to have the mass of the Sun with the Galactic abundance ratios of elements, and $A_{\rm min} = 16$ gives a fractional abundance of OS and heavier elements of $8 \times 10^{-9}$; if only heavier elements bind then the abundances increase correspondingly.   Clearly, a much more detailed calculation is needed taking into account the evolution of elemental abundances and the diversity in sizes and ages of the stars contributing to the pre-solar material, but this back-of-envelope estimate indicates that there may be a significant pre-solar contribution to the local exotic isotopes.  If this pre-solar component is dominant, that would be evidenced by all samples for a given element having a common abundance fraction, whether of a lunar, ocean, mantle or other terrestrial origin.  In principle, evidence for such exotic isotopes could appear as extra lines in the vibrational spectra of cold molecular clouds, although given the very low abundance levels it may not be possible.

\subsection{Exotic isotope detection} 
 \label{sec:hybDet}
 
The mass of the exotic isotope is
\be
\label{massAS}
M_{AS} = 
M_{A} + m_S - BE_A(\alpha) ~,
\ee
where $BE_A(\alpha)$ is the ground-state binding energy.   The main uncertainty in predicting $M_{AS} $ is in the value of $m_S$.  Based on the discussion in earlier sections,  $m_S$ can be anticipated to be in the range 1800-2000 MeV.  For our fiducial value $m_S \approx 2 \, m_p$, the exotic isotope is similar in mass to the $A+2$ isotope.   The binding energy is shown in Fig.~\ref{fig:BE} (henceforth replacing $\alpha_{SN} \rightarrow \alpha$ for brevity).  
Generally, $BE_A(\alpha)$ ranges from 0 to $\approx 10$ MeV.  

However isotopes of the ``wrong" mass may well have gone unnoticed.  For instance, $^{18}$O is a stable isotope which accounts for 0.2\% of naturally occurring oxygen.   If one in $10^9$ of  $^{16}$O captures an \sdm, that would produce an isotopic mass similar to naturally occurring $^{18}$O at a ppm level.   Similarly the most abundant isotope of Si is $^{28}$Si, but 3\% of Si is $^{30}$Si, potentially masking detection of the anomalous isotopic mass.
Modern mass spectroscopy tools used for geochemical research have the sensitivity and resolution to discover $OS$ at the predicted level, with 
mass resolution in the  $A\approx 16$ regime of at least 1 part in 30,000 ($\approx 0.5\,$MeV) and up to 1 part in 1 million for certain specialized applications~\cite{Eiler+MassSpec13,Eiler+MolecIso17}.   
For smaller \asn\ the threshold $A_{\rm min}$ for forming exotic isotopes increases, and estimating the abundance of different elements is more complex due to the non-trivial geochemistry of the Earth's crust.  On the other hand, heavier nuclei are rarer so the fraction which are exotic would be higher, so robust limits on the entire range \asn\ $\gtrsim 0.04$ may be possible through dedicated searches.  

Although predicting the abundances of different exotic isotopes for general \asn\ is challenging, the absence of  comprehensive and accurate fractional abundance predictions does not preclude an effective experimental search.  A potential signal is a narrow line in a sensitive mass spectrometer at a mass not known to be associated with known isotopes.  Let us denote the mass of the unexplained line as $\mu_Z$.  Initial searches would focus on mass ranges $\mu_Z  = M_{Z,A_i} \pm (1800-2000)\,$MeV, where $i$ denotes a stable isotope of $Z$.  There are a number of signatures to discriminate a real sexaquark bound state signal from noise or other sources.  If it is a bound state of $\{Z,A\}$ and sexaquark, then 
\be
\mu_Z = M_{Z,A} + m_S - BE_A(\alpha)~.
\ee
There is no ambiguity as to which $\{Z,A_i\}$ could be the bound state host nucleus because the range of mass uncertainty, 100 MeV, $<<$ 1 amu.   

Three complementary classes of consistency conditions are:\\

\noindent $\bullet$   \underline{Multiple stable isotopes $\{Z, A_i\}$ }\vspace{0.06in}\\
Each isotope of the same element should have a similar relative abundance of \sdm\ bound states, to the extent that their geochemical history is the same and $\sigma^{\rm cap}$ varies slowly with $A$.  Thus a line of similar strength should be found associated with each stable isotope, with the line positions satisfying
\be
\label{Deltaij}
\mu_{Z,i} - \mu_{Z,j}  = M_{Z,A_i} - M_{Z,A_j} - BE_{A,i}(\alpha) - BE_{A,j}(\alpha)~.
\ee
$BE$ is a slowly varying function of $A$ \cite{xf20}, so the splitting in Eq. (\ref{Deltaij}) is generically small, $\lesssim$ 1 MeV.  To the extent that the exotic abundances are the same, the line strengths relative to those of the unbound isotopes should be similar.\\

\noindent $\bullet$   \underline{ Hosts with neighboring $\{Z,Z'\}$}\vspace{0.06in}\\
If elements can be studied which have neighboring $\{Z,Z'\}$ with both having stable isotopes with the same $A$, the prediction is even stronger:
\be
\mu(Z,A) \approx \mu(Z',A)
\ee 
with the equality only broken due to non-identical nuclear wavefunctions of $\{Z,A\}$ and $\{Z',A\}$.   Here, however, the fractional abundances and hence signal strength are not in general the same, since the geophysics and geochemistry of the two elements are different.\\
 
\noindent $\bullet$   \underline{Multiple observed lines for different $\{Z,A\}$ hosts}\vspace{0.06in}\\
With signals seen for $N \geq 3$ different $\{Z,A\}$ hosts, one can extract $m_S - BE(A)$ for each of them.  Within the uncertainty on $BE_A(\alpha)$ due to uncertainties in the nuclear wavefunctions, a single value of $m_S$ and $\alpha$ should give a good fit to all $\mu_{Z,A}$.

\section{Accelerator Searches for the Sexaquark}
\label{sec:searches}

One of us (GRF, ~\cite{fS17}) made a comprehensive review of experimental searches for the H-dibaryon and determined that apart from one experiment, all of the searches up to 2017 either searched for decay products or were only sensitive to masses below 2 GeV, making them insensitive to the hypothesized stable \s.  The one exception,  BNL E888 ~\cite{belz+DiffDissocH96}, placed bounds that were far too weak to be relevant to its existence \cite{fS17}.  The experimental challenge to discovering an \s\ and demonstrating its existence is that \s's are similar in mass to neutrons, but interact less and are much less abundant.   In a low energy exclusive reaction such as $K^- p \rightarrow  S \bar{\Lambda}$, the quantum numbers of the unseen \s\ are unambiguous, but the rate \s\ production $ \sim \tilde{g}^2$.   If the DM is composed of sexaquarks, $\tilde{g}^2\lesssim 4 \times 10^{-12}$ (see Sec.~\ref{sec:durability}), and the expected suppression is even more severe if the theoretical estimate for $\tilde{g}$ (Sec. \ref{sec:buTh}) is used.  

The situation in a high energy collision is different, and we must consider two regimes.  In an inclusive short-distance-initiated reaction like decay of Upsilon(1S,2S,3S), closure approximation is valid and estimation is relatively robust.  The analysis is given in \cite{fS17} and recapped in Supplemental Materials \ref{app:Penalty}.  Estimating the \s:n production ratio in the central region of high energy collision is more uncertain.  One appraoch is to use the naive rule of thumb based on baryons versus mesons that for each additional quark which has to be incorporated into a hadron, the abundance is reduced by a factor 10-20.  Producing an \s\ entails 3 additional quarks that must be incorporated into the state, for a penalty of a factor $10^3$, and baryon number conservation requires the production of still another anti-baryon relative to the case of neutron production, for an additional penalty of $10^{1-3}$ depending on how that should be counted.  Since no penalty is imposed on account of the \s's small size this may be an overly optimistic estimate, but it suggests that in a very high energy collision or in the final state of $Z$ decay, the abundance of \s's (or \sbar's) could be $10^{-4} - 10^{-6}$ relative to neutrons.    

However it is not enough to produce an \s\ -- it must be identified, or its presence unambiguously established, which is non-trivial.  Unlike in a search for heavy Beyond the Standard Model neutral particles, e.g., as expected in Supersymmetry \cite{f:23}, the \s\ has a typical QCD transverse momentum, $\mathcal{O}(1)$ GeV, so the missing energy or missing transverse momentum it carries is too small to notice at a high-energy accelerator.   Two methods to search were suggested in \cite{fS17}: searching the final states of $\Upsilon$ decay and searching for a very distinctive signature of \sbar\ annihilation at the LHC.   They are briefly recalled below along with several new suggestions.\\

\noindent $\bullet$ \underline{Final states of $\Upsilon(1S,2S,3S)$ decays}

The reactions
\be
\label{ups}
\Upsilon~~[ \rightarrow {\rm gluons}] \rightarrow S \, \bar{\Lambda} \, \bar{\Lambda}~~{\rm or} ~\bar{S} \, \Lambda \, \Lambda~~ + {\rm pions~ and/or} ~\gamma 
\ee
are ideal discovery channels.   The characteristic size of the $ggg$ state from which the final hadrons emerge is (10 GeV)$^{-1} = 0.02 $ fm, so the small size of the \s\ is not an inherent disadvantage, as it is when produced in a reaction like $K^- p \rightarrow  S \bar{\Lambda}$.   A statistical estimate of the branching fraction for {\it inclusive} \s\ plus \sbar\ production is $ 2.7 \times 10^{-7}$;  see the Supplemental Materials for details.  It is small due to the necessity of producing extra gluons to have the required minimum 6 $q \bar{q}$ pairs, and the low probability of 6 quarks or antiquarks of the required flavors being in a color-flavor-spin singlet state.   

If all of the final particles are seen, the mass of the unseen \s\  can be reconstructed from 4-momentum conservation:  $m_S^2 = (p_\Upsilon - p_{\Lambda 1}- p_{\Lambda 2} - \Sigma p_{\pi's\&\gamma})^2$.  The width of the missing-mass peak is entirely due to resolution which is so good in some detectors, $\mathcal{O}(20)$ MeV, that even a few events appearing to be $\bar{\Lambda} \bar{\Lambda}$ or $\Lambda \Lambda $ + pions or gamma, having a common missing mass, would be a powerful smoking gun for the existence  of the \s\, and would accurately determine its mass.   
The initial state can be any $\Upsilon$ or continuum state below open-bottom threshold.  Other final states besides $\Lambda \Lambda$/$\bar{\Lambda} \bar{\Lambda}$ are also discovery avenues, e.g., $\Xi^- p$, or a $\Lambda $ can be replaced by $K^- p $.  As long as no B- and S-carrying particle escapes detection besides the \s\ or \sbar, any combination of hyperons and mesons with B= $\pm$ 2, S= $\mp$ 2 quantum numbers, including final states with higher multiplicities, can be used.   
The $\bar{\Lambda} \bar{\Lambda}$ and $\Lambda \Lambda $ final states are very good because the $\Lambda $'s short decay length ($c \tau = 8$ cm) and 64\% branching fraction to the 2-body charged final state $p \, \pi^- $, means $\Lambda$'s and $ \bar{\Lambda}$'s can be reconstructed with high efficiency, and their 4-momenta well-measured.  

Babar performed a search for {\it exclusive} \s\ and \sbar\ production and placed an upper limit on the branching fraction of 
BF$_{\rm exc} < 1-2 \times 10^{-7}$ \cite{babarS18}.  However as is shown in the Supplemental Materials by examining exclusive branching fractions for other channels in $\Upsilon$ decay, the penalty for demanding an exclusive final state is at least a factor $10^4$.  Thus Babar's sensitivity in the search \cite{babarS18},  is by far insufficient to shed light on the possible existence of a stable \s.  (The utility of $\Upsilon$ decay final states as being potentially enriched in flavor-singlets, in particular the H-dibaryon, was recognized early-on by Belle, however their search~\cite{BelleHdibaryon13} assumed the H-dibaryon was unstable and sought evidence of its decay into $\Lambda$ final states, so is inapplicable to the sexaquark scenario.)

A more general strategy than just identifying events with exactly two $\Lambda$s or two $\bar{\Lambda}$s plus only pions or gammas in final states of $\Upsilon(1S,2S,3S)$ decay, is to study the proportion of events having specified numbers $\{N_B, N_S, N_{\bar{B}},N_{\bar{S}}\}$ of baryons, strangeness +1 particles, anti-baryons and strangeness -1 particles, respectively, in the final state.  The feasibility of establishing a statistically significant excess of events with the correlated $N_B - N_S = \pm 4$ expected in the case of \s\ and \sbar\ production, depends on $N_{\rm tot}$, the total number of $\Upsilon(1S,2S,3S)$ decays recorded, and the ID efficiency of the various baryons and strange particles, including losses from less than $4 \pi$ detector coverage.  In the Supplemental Materials a simple estimate is given, using a single effective efficiency for identifying baryons and anti-baryons, $e_{B}$, and similarly $e_{S}$ for strange and anti-strange particles, to roughly assess the feasibility.  It suggests that the estimated inclusive branching fraction $2.7 \times 10^{-7}$ may be accessible to Belle-II, depending on the amount of running on $\Upsilon(1S,2S,3S)$ and the actual effective efficiencies.  This motivates a more detailed investigation with a real detector simulation.  The hadronic event generator EPOS-LHC has been modified to incorporate \s\ and \sbar\ production in hadron and heavy ion collisions, and in $\Upsilon(1S,2S,3S)$ decay, with a coalescence model production mechanism \cite{EPOS-S20}.
 \\

\noindent $\bullet$ \underline{\sbar\ annihilation in an LHC tracker}

With some rate, perhaps $10^{-4} - 10^{-6}$ relative to neutrons according to the simplistic estimate above, \sbar's should be produced in LHC collisions.  In a detector such as CMS, ATLAS, ALICE or LHCb, such an \sbar\ can annihilate with a nucleon in the material of the beam-pipe or tracker, to produce a very distinctive final state in which for instance a $\bar{\Lambda}$ and a K emerges from the material.  Unfortunately, the expected rate of useful events is very small.  If the lab energy of the \sbar\ is small enough for the annihilation event to not produce too many final particles, the cross section  $\sim \tilde{g}^2$, i.e., very small.  If the energy is large and many particles are produced, possibly evading the breakup amplitude suppression, then identifying the characteristic B  =  -1 and S = +2 signature of the final state is almost hopeless.  See \cite{fS17} for more detailed discussion.
\\

\noindent $\bullet$ \underline{Search for long-interaction-length stable neutral particle}

The challenge in searching for inclusive \s\ and \sbar\ production in a high energy collision is the problem of identifying them in the face of vastly more neutrons, as mentioned earlier.  A possible strategy is to search for evidence of a neutral component with interaction length longer than that of neutrons and different from known neutral long-lived particles.  Due to the small value of $\tilde{g}$, the \sbar\ annihilation channel is much smaller than its scattering channel, so \sbar\ interactions should be very similar to \s\ interactions.   The \s\ and \sbar's  are generally relativistic even in the central region, so the calculations in Sec.~\ref{sec:DMints} do not apply.  Instead, we can roughly estimate their interaction length relative to that of neutrons in this energy regime as $\lambda^{\rm int}_S \approx (\alpha_{NN}/\alpha_{SN})^2 \,\lambda^{\rm int}_n \approx 6 \times 10^3 \,(0.2/\alpha_{SN})^2\, \lambda^{\rm int}_n$, with coupling $\alpha_{SN}$ taken to be the same to first approximation as the $\alpha_{SN}$ which enters the potential scattering problem relevant for dark matter constrained in Sec.~\ref{sec:DMints}.

The strategy of looking for an anomalous component of long-interaction-length neutral stable particles could be implemented with a relatively simple customized experiment. 
Conceptually, a beam is directed onto a target, followed by sweeping magnets and decay region to eliminate charged particles and short-lived neutral components.  This would be followed by an instrumented region with particle tracking interleaved with absorber, whose overall length is $\gtrsim 10^5$ neutron interaction lengths. The requirements on the tracking being to measure the longitudinal position of n- and \s- or \sbar-initiated events, and to discriminate between interactions and decays, which need to be rejected.   A quasi-beam-dump setup could be employed to reduce the $n:S$ ratio in the detector.   The detector could be built up in stages, initially adapted to a shorter interaction length in case \asn\ is larger than $0.2$.   An in-principle-complementary approach is exemplified by the experiment of \cite{gustafson} at Fermilab which used timing rather than anomalous interaction length.  It was only sensitive to masses above 2 GeV due to neutron background, so not applicable for the sexaquark.  However the method requires measuring the energy deposit and time-of-flight to discriminate the new particle from neutrons so seems to be both more complicated and less powerful -- but perhaps merits consideration.\\

\noindent $\bullet$ \underline{Heavy Ion Collisions}

A very attractive production channel for \s\ and \sbar\ is in the central region of relativistic heavy ion collisions (HICs), because of the similarity of the production to that of DM in the Early Universe.  The process is not identical to the Early Universe because in the Early Universe the cooling timescale at the hadronization transition is $\sim 10^{-5}$s and the medium is infinite, while in a heavy ion collision the cooling time is very much shorter and the plasma expands into the vacuum.

Ref. \cite{Andronic+HIC_Nature18} obtains an excellent fit to the relative abundances of final particles in central Pb-Pb collisions, including such complex and exotic states as hyper-triton, assuming statistical equilibrium at a temperature $T=156$ MeV and accounting for production and decays of resonances.   The main systematic uncertainty is associated with treatment of the resonances.  A similar approach applied to \s\ and \sbar\ production would give a result similar to deuteron and anti-deuteron:  $dN/dY \approx 10^{-1}$ in the central Pb-Pb collisions at $\sqrt{s_{NN}} = 2.76$ TeV -- about a factor-300 less than $p$ and $ \bar{p}$. 

Perhaps the long-interaction-length neutral particle technique can be employed, depending on the particulars of the detector.  Another strategy is to look for an excess of events in which the observed final state has  baryon number minus strangeness $ | \, B - S | = 4$ , due to production and escape of an \s\ or \sbar\ whose baryon number and strangeness is balanced by the observed final state hadrons.  The problem of course is the impossibility of perfectly measuring the B and S of each final particle.  A further problem for ALICE is the limited rapidity range that can be observed.  In Ref. \cite{ALICE_Bflucs_19}, ALICE presents a study of the event-to-event fluctuations in the baryon number of particles with $0.6 < p < 1.5$ GeV/c and $| \eta | < 0.8$.  For central collisions the difference in number of baryons and anti-baryons is of order the sum.  With such large fluctuations, it would appear difficult to discern a population of events above background with $B-S = \pm 4$, unless a large portion of the final particles can be ID'd.  A detector simulation or an analysis along the lines of Supplemental Materials \ref{app:B-S=4} would be needed to properly assess the prospects.
\\

\noindent $\bullet$ \underline{High intensity photon beams}

J-Lab has a tagged photon beam of energy 9 to 12 GeV, with $10^8$ photons/second on target.  The GlueX experiment anticipates collecting  $\approx 10^{12}$ interactions.  It has adequate kinematic reach to probe reactions such 
\be
 \gamma \, p \rightarrow S \, \bar{\Lambda} \, K^+ + pions ~.
 \ee  
A 12 GeV photon provides $E_{\rm CM} = 4.84$ GeV; this is 1.35 GeV above the 3.5 GeV total mass of $S, \, \bar{\Lambda}, \, K^+$ for the fiducial $m_S = 2 m_p$, leaving room to spare for phase space and pion production.   Depending on the solid angle coverage and tagging efficiency, unbalanced baryon number and strangeness due to an escaping \s\ could be a good discovery channel in spite of the \s\ production rate going as $\sim \tilde{g}^2$, given the potentially very large number of events.

\section{Summary}
\label{sec:Sum}

At this point in time, it is not possible to decide on theoretical grounds whether there is an as-yet-undiscovered stable neutral boson with baryon number 2 in the QCD spectrum.  Many would argue against this possibility based on faith in the qualitative and semi-quantitative understanding the community has developed of chiral symmetry breaking, quark confinement, asymptotic freedom and many aspects of the hadron spectrum in QCD.  However in the spirit that our understanding of QCD may not be as complete as we would like to think, and motivated by the lack of a compelling and viable alternative Dark Matter candidate, we have examined here whether a stable sexaquark can be excluded in light of present knowledge, and whether it would make a satisfactory DM particle.

We find that sexaquark dark matter is not presently excluded either by accelerator experiments, dark matter direct detection constraints, astrophysical constraints such as neutron stars and SN1987a, or cosmological constraints such as the CMB and structure formation or primordial nucleosynthesis. 

The sexaquark relic abundance after the transition from the quark gluon plasma to the hadronic phase is completely determined in freezeout approximation by statistical physics, known parameters of QCD and the sexaquark mass.  The predicted value is $\Omega_{DM}/\Omega_b \approx 5$, with better than factor-2 accuracy.   This in remarkable agreement with the observed value $\Omega_{DM}/\Omega_b = 5.3 \pm 0.1$.   The analysis is given in Sec. \ref{sec:DMabun}.  

Whether this no-free-parameter result persists to low-temperatures depends on the amplitude to break up a sexaquark into two baryons, \gsbb, which we calculate in Sec.~\ref{sec:buTh}.  Three separate effects combine to suppress sexaquark breakup, the most important being the hard-core radius characterizing the short-distance repulsion of the baryon-baryon potential;  we adopt the central value of the standard fits to data, $r_c = 0.4$ fm~\cite{HamadaJohnston62,Reid68}.  Other sources of suppression are the small radius of the sexaquark and the QCD barrier to tunneling through the intermediate state.  Predictions for \gsbb\ for extreme choices of the parameters are shown in Fig. \ref{fig:gsbb}, where one sees that even with no tunneling suppression and taking the maximum $r_S$, the breakup amplitude is too small to destroy the abundance ratio set at the QGP-hadron transition.

We also constrained the allowed range of \gsbb\ empirically, most importantly by using SNO data to obtain a limit on the deuteron beta decay lifetime: $\tau_D > 10^{29}$ yr.  This gives the strongest and most robust experimental limit to date on the sexaquark breakup amplitude if the \s\ is lighter than 1870 MeV.  It improves on and superceeds the estimates derived in \cite{fzNuc03} for Oxygen decay in SuperK.  The limits on \gsbb\ are summarized in Fig. \ref{fig:gsbb}.  The mass range 1870-1880 MeV is virtually unconstrained, but any mass above 1850 is currently comfortably compatible with the constraints, given theoretical estimates for the breakup amplitude.

Since the \s\ is a particle in QCD, its scattering on baryons can be related to parameters of effective field theory.  The scattering is primarily due to exchange of the flavor singlet linear combination of $\omega$ and $\phi$ vector mesons, here denoted $V$ (Sec. \ref{sec:DMints}).  Thus \sdm\ interacts with baryons via a Yukawa potential with range $m_V^{-1} \approx 0.2 $ fm.  The main uncertainty in the calculation of DM-nucleus cross sections is the strength of the potential, \asn, and its sign.  At leading order this is determined by the coupling between \s\ and $V$, $g_{SSV}$, with $| \alpha_{SN} | = | g_{SSV} \, g_{NNV} | /(4 \pi)$; we consider $0.001 < \alpha_{SN} < 10$.  

We developed the necessary theoretical infrastructure to interpret DM direct detection experiments when the DM has non-perturbative interactions with nucleons.  Cross sections do not scale with atomic mass $A$ in the manner widely assumed based on Born approximation, and the form factor for extended nuclei and the velocity dependence of DM-nucleus cross sections exhibit strong, highly non-trivial behavior.  We treated these in detail by solving the Schroedinger equation.  We obtained constraints on the Yukawa coupling parameter \asn\  as a function of $m_S$ implied by direct detection experiments and constraints on $\sigma_{\rm DM-p}$ from  the CMB power spectrum.  

The excluded regions for an attractive interaction are shown in Fig. \ref{fig:DDexclusions}; the boundaries in the repulsive case are smooth but similar.  For a repulsive interaction and $m_S \approx 2 m_p$, the maximum Yukawa coupling is $| \alpha_{SN} |_{\rm max} \approx 0.004$ and $\sigma_{SN} \lesssim 10^{-29} {\rm cm}^2$, taking the dewar limits~\cite{nfm18,nbn19} at face value; see Sec.~\ref{sec:dewar} for a discussion of a more conservative choice.

For an attractive interaction, $ \alpha_{SN}$ can be as large as 0.7 and the phenomenology of \sdm\ is very rich.  The constraint $\alpha_{SN} < ~\approx 0.7$ is required to avoid that virtually all primordial $^4$He is in the form of a bound state with an \s\ whose mass is about 2 amu larger than normal $^4$He.  But for $\approx 0.7 > \alpha_{SN} \gtrsim 0.03$,  \sdm\ captured by Earth quickly binds to nuclei.  For example, if $\alpha_{SN} \gtrsim 0.2$, \sdm\ binds to oxygen nuclei in the Earth's oceans, crust and atmosphere, forming a roughly 100-part-per-billion abundance of exotic isotopes of  $O$ with a mass about 2 amu larger than the fundamental nuclear mass.   Experimental searches for exotic isotopes have not explored such small mass offsets with adequate (ppb) sensitivity.  Future searches of this type are well-motivated because $\alpha_{SN} \gtrsim 0.004$ seems natural so the dewar limits for the repulsive case suggest the \s-baryon interaction is attractive.  

\section{Conclusions}
\label{sec:Conc}

We have shown that sexaquarks are an excellent dark matter candidate.   Their relic density is fixed by the physics of the transition from quark gluon plasma to hadrons and statistical physics and known parameters from QCD predicts the DM to baryon density ratio \ODMOb\ $\approx 5$, in excellent agreement with the observed value \ODMOb = $5.3 \pm 0.1$.   

The possible range of sexaquark coupling strength to baryons considered here, $0.001 \lesssim \alpha_{SN} \lesssim 10$, allows for a broad range of phenomenological behaviors.  The \sdm-proton scattering cross section may be $\sigma_{S{\rm p}} \lesssim 10^{-29}\,{\rm cm}^2$, which would have evaded detection and is compatible with all cosmological, astrophysical and laboratory bounds;  for a repulsive \sdm-baryon interaction this is the only option compatible with observational limits.  However the DM-nucleon interaction is {\it a priori} equally likely to be attractive, in which case the phenomenological options are much richer.  

The sexaquark dark matter hypothesis motivates the following experimental efforts:
\\
$\bullet$ Search for exotic isotopes of $A>4$ nuclei in the Earth, having a mass offset of $m_S - BE$, with $BE \lesssim 10 \,{\rm MeV}$, i.e., about 2 amu heavier than the host nucleus $A$.  If the Yukawa coupling between \sdm\ and nucleons is attractive and strong enough to bind to oxygen, $\gtrsim \mathcal{O}$(ppb) of oxygen nuclei should be exotic and have a mass within 10's or at most 100's of MeV of $^{18}$O.  For weaker Yukawa coupling only heavier nuclei bind, motivating the search for exotic isotopes of a diversity of elements.  If DM proves to be composed of sexaquarks which can bind to nuclei, the study of the abundances of exotic isotopes will become a powerful tool for geoscience as well as a high-precision window on the sexaquark mass and its coupling to nucleons.
\\
$\bullet$ Search in final states of $\Upsilon(1S,2S,3S)$ decay, and also in heavy ion collisions and at J-Lab, for the signature of \s\ or \sbar\ production.    In Upsilon decay one can search for events with $\bar{\Lambda}\bar{\Lambda}$ or $\Lambda \Lambda$ and no accompanying baryon number or strangeness.  Even a handful of events in which all particles except the unseen \s\ or \sbar\ are  well-measured, could give a missing-mass peak that is a decisive signature.  Another strategy would be an inclusive search for events with $B-S = \pm 4$. 
\\
$\bullet$ Search for the presence of a second component in the interaction-length distribution of stable neutral particles produced in the central region of relativistic heavy ion or other high energy collisions.   Taking anti-deuterium production as a guide, the abundance of the \s\ and \sbar\ component in a heavy ion collision should be about 0.3\% that of $n$ and $\bar{n}$.  The anomalous interaction length depends on \asn\ which is constrained by cosmological and dewar constraints as discussed in Sec.~\ref{sec:DMints}; a rough estimate is $\lambda^{\rm int}_S \approx 6 \times 10^3\, \lambda^{\rm int}_n \, (\alpha_{SN}/0.2)^2$.
\\
$\bullet$ Calibrate the XQC and other semi-conductor detectors used to search for DM with mass less than a few GeV, to determine the extent to which low energy recoiling atoms deposit their energy in forming lattice point defects (Frenkel pairs) or coherent phonon excitations (thermalization).  If the fraction of energy thermalized is large enough, detectors such as XQC above the atmosphere or CRESST near the surface of Earth would be sensitive to DM mass in the $\approx 2 m_p$ range.  However until these detectors are calibrated for low recoil energies, they can only be used to place limits on DM interactions above the mass range relevant for \sdm.  



\begin{acknowledgments}
Over the course of this research, we have benefitted from helpful discussions and input from many colleagues, including Y. Ali-Haimoud, D. Blaschke, P. Braun-Munzinger, F. Buccella, J. Carlson, S. Dubovsky, B. Echenard, J. Eiler, R. Galvez, A. Haas, I. Jaegle, S. Lowette, M. S. Mahdawi, C. McKee, P. Molaro, R. Mussa, D. Neufeld, S. Olsen, M. Pospelov, T. Rijken, J. Schaffner-Bielich, J. Ruderman, M. Unger, D. Wadekar, N. Wintergerst and R. Wiringa.
The research of GRF was supported in part by the Simons Foundation and NSF-1517319; XX received support from a James Arthur Graduate Fellowship.

\end{acknowledgments}

\providecommand{\noopsort}[1]{}\providecommand{\singleletter}[1]{#1}%

\newpage

\end{document}